\documentclass[aps,twocolumn,showpacs]{revtex4}
\usepackage{amsmath,color,ulem}
\usepackage{multirow}
\usepackage{graphicx}
\usepackage{graphicx,color,dcolumn,booktabs,bm,multirow}
\usepackage{longtable,lscape}
\begin{document}

\title{Search for singly charmed dibaryons in baryon-baryon scattering}

\author{Yao Cui$^1$}\email{cuiyao.990213@163.com}
\author{Xinmei Zhu$^2$}\email{xmzhu@yzu.edu.cn}
\author{Yuheng Wu$^3$}\email{Wuyuheng@ycit.edu.cn}
\author{Hongxia Huang$^1$}\email{hxhuang@njnu.edu.cn(Corresponding author)}
\author{Jialun Ping$^1$}\email{jlping@njnu.edu.cn}
\affiliation{$^1$School of Physics and Technology, Nanjing Normal University, Nanjing 210097, People's Republic of China}
\affiliation{$^2$Department of Physics, Yangzhou University, Yangzhou 225009, People's Republic of China}
\affiliation{$^3$Department of Physics, Yancheng Institute of Technology, Yancheng 224000, People's Republic of China}

\begin{abstract}
 We perform a systematical investigation of the singly charmed dibaryon system with strangeness numbers $S=-1$, $-3$ and $-5$ in the framework of the chiral quark model.
 Two resonance states with strangeness numbers $S=-1$ are obtained in the baryon-baryon scattering process. In the $\Lambda\Lambda_{c}$ scattering phase shifts, the $\Sigma\Sigma_{c}$ appears as a resonance state with the mass and width 3591 MeV and 11.1 MeV, respectively. In the $N\Xi_{c}$ and $N\Xi^{\prime}_{c}$ scattering phase shifts, the $\Sigma\Sigma^{\ast}_{c}$ exhibits as a resonance state with the mass and width 3621-3624 MeV and 14.9 MeV, respectively.  All these heavy-flavor dibaryons are worth searching for in experiments. Besides, we would like to emphasize that the coupling calculation
between the bound channels and open channels is indispensable. The study of the scattering process maybe an effective way to look for
the genuine resonances.
\end{abstract}

\maketitle

\setcounter{totalnumber}{5}

\section{\label{sec:introduction}Introduction}
In the last two decades, a growing number of exotic particles have been discovered in experiment. A series of $XYZ$ states, $P_{c}$ states, and charm $T_{cc}^{+}$ state were reported in experiment, which has led to extensive research into exotic hadrons ~\cite{XYZ1, XYZ2, XYZ3, XYZ4}. Understanding hadron-hadron interactions and searching for exotic hadron states are important topics in hadron physics, among which questing for dibaryons is a long-standing challenge.  The well-known dibaryon is deuteron discovered in 1932 ~\cite{deutron}. In 2014, the Wide Angle Shower Apparatus (WASA) detector at the Cooler Synchrotron (COSY)~\cite{WASA1, WASA2} collaboration established the narrow resonance state $d^{\ast}$ with $I(J^{P})=0(3^{+})$, and given the first clear-cut experimental evidence for the existence of a true dibaryon resonance~\cite{WASA3}. The $d^{\ast}$ (2380) may be a $\Delta\Delta$ dibaryon state or a six-quark state, and extensively investigated within various theoretical approaches~\cite{dstar1, dstar2, dstar3, dstar4}.

For the strange dibaryon, the progress of the $N\Omega$ searches in the experiment attracted more and more attention for this state, which was observed in Au+Au collisions by STAR experiment at the Relativistic Heavy Ion Collider (RHIC)~\cite{nomega1}. And before that, the dibaryon $N\Omega$ was investigated by different theoretical methods such as quark models~\cite{nomega2, nomega3, nomega4, nomega5, nomega6}, and the lattice QCD~\cite{nomega7, nomega8}.

The research of charmed dibaryon is further inspired by the experimental discovery of the doubly charmed baryon $\Xi_{cc}$ by the Large Hadron Collider beauty (LHCb) Collaboration~\cite{LHCb}. For the dibaryons with heavy quarks, the $N\Lambda_{c}$ system with one heavy quark was both studied on the hadron level~\cite{c1} and on the quark level~\cite{c2}. The dibaryon systems with two heavy quarks were researched in the one-pion-exchange model~\cite{cc1} and one-boson-exchange model~\cite{cc2, cc3}. Besides, the dibaryon systems with three heavy quarks were also investigated from the lattice QCD~\cite{ccc1}, the QCD sum rule~\cite{ccc2}, one-boson-exchange~\cite{ccc3, ccc5} and the quark model~\cite{ccc4}. Recently, Junnarkar and Mathur reported the first lattice QCD study of the heavy quark flavor deuteron-like dibaryons~\cite{ccc1}, and suggested that the dibaryons $\Omega_{c}\Omega_{cc}(sscscc)$, $\Omega_{b}\Omega_{bb}(ssbsbb)$ and $\Omega_{ccb}\Omega_{cbb}(ccbcbb)$ were stable under strong and electromagnetic interactions. They also found that the binding of these dibaryons became stronger as they became heavier in mass. In addition, there are many other investigations on deuteron-like states. In Ref.\cite{cc2}, they perform a systematic study of the possible loosely bound states composed of two charmed baryons or a charmed baryon and an anticharmed baryon within the framework of the the one-boson exchange model. And in Ref.\cite{ccc5}, they also adopted the one-boson-exchange model to perform a systematic investigation of interactions between a doubly charmed baryon ($\Xi_{cc}$) and a $S$-wave charmed baryon ($\Lambda_{c}, \Sigma_{c}^{\ast}$ and $\Xi_{c}^{\prime,\ast}$), which can be easily bound together to form shallow molecular hexaquarks. Taking inspiration from the research on the dibaryon states containing heavy quarks, it is meaningful to use various methods to study and search for these heavy dibaryons.

 Quantum chromodynamics (QCD) is a theory describing strong interactions based on regular field theory. The equivalent degrees of freedom are quarks and gluons, and QCD is asymptotically free at high energies and can be solved precisely by perturbation theory. Generally, hadronic structure and hadron interactions belong to the low-energy physics of QCD, which are much harder to calculate directly from QCD because of the nonperturbative nature of QCD. One must rely on effective theories or models inspired by QCD to gain insight into the phenomena of the hadronic world. The constituent quark model is one of them, which transforms the complicated interactions between current quarks into dynamic properties of constituent quarks. The chiral quark model (ChQM) is a typical one of the constituent quark model. The ChQM was successfully used to calculate mesons~\cite{2, meson}, baryons, tetraquarks~\cite{4, chqm2}, pentaquarks~\cite{5} and dibaryons~\cite{nomega5}. In particular, for dibaryon systems, the ChQM is able to calculate the dibaryon systems from light to heavy quarks very well, such as nucleon-nucleon interaction~\cite{nn}, $N\Omega$~\cite{nomega5} and the fully heavy dibaryon systems~\cite{full}, which is consistent with the results of the lattice QCD.

In the present work, we systematically investigate the singly charmed dibaryons in the ChQM, where the effective potential between two baryons are evaluated, and the search of possible bound states are performed with the coupled channel effects. Moreover, based on the conservation of the quantum numbers and the limitation of the phase space, we also study the baryon-baryon scattering process to look for the existence of any resonance states in the singly charmed dibaryon systems.

The structure of this paper is as follows. A brief introduction
of the quark model and calculation methods are given in Section II. Section III is devoted
to the numerical results and discussions. Section \uppercase\expandafter{\romannumeral4} is a
summary and the last section is Appendix, which shows the way of constructing wave functions.

\section{Quark model and calculation methods}
Phenomenological model is an important tool to analyze the nature of multi-quark states. Here, the chiral quark model(ChQM) is used to study the singly charmed dibaryon systems with $IJ=01$. In addition, the six-body problem is transformed into a two-body problem by using the resonance group method(RGM) for simplified calculations.

\subsection{The chiral quark model}
The model has become one of the most common approaches to describe hadron spectra, hadron-hadron interactions and multiquark states~\cite{ChQM(RPP)}. The construction of the ChQM is based on the breaking of chiral symmetry dynamics~\cite{chqm1, chqm2}. The model mainly uses one-gluon-exchange potential to describe the short-range interactions, a $\sigma$ meson exchange (only between u, d quarks) potential to provide the mid-range attractions, and Goldstone boson exchange potential for the long-range effects~\cite{chqm3}. In addition to the Goldstone bosons exchange, there are additional $D$ meson that can be exchanged between u/d and c quarks, $D_{s}$  meson that can be exchanged between s and c quarks, and $\eta_{c}$ that can be exchanged between any two quarks of the u, d, s and c quarks. In order to incorporate the charm quark well and study the effect of the $D$, $D_{s}, \eta_{c}$ meson exchange interaction, we extend the model to $SU(4)$, and add the interaction of these heavy mesons interactions. The extension is made in the spirit of the phenomenological approach of Refs. ~\cite{Glozman, Stancu}.
The detail of ChQM used in the present work can be found in the references~\cite{ChQM1,ChQM2,ChQM3}. In the following, only the Hamiltonian and parameters are given.

\begin{widetext}
\begin{eqnarray}
H &=& \sum_{i=1}^6 \left(m_i+\frac{p_i^2}{2m_i}\right) -T_c
+\sum_{i<j} \left[ V^{CON}(r_{ij})+V^{OGE}(r_{ij}) + V^{\sigma}(r_{ij})+ V^{OBE}(r_{ij})
\right], \\
V^{CON}(r_{ij})&=& -a_c {\boldsymbol \lambda}_i \cdot {\boldsymbol
\lambda}_j [r_{ij}^{2}+V_0],  \\
V^{OGE}(r_{ij})&=& \frac{1}{4}\alpha_{s} \boldsymbol{\lambda}_i \cdot
\boldsymbol{\lambda}_j
\left[ \frac{1}{r_{ij}}-\frac{\pi}{2}\left(\frac{1}{m_{i}^{2}}
 +\frac{1}{m_{j}^{2}}+\frac{4\boldsymbol{\sigma}_i\cdot
 \boldsymbol{\sigma}_j}{3m_{i}m_{j}}  \right)
 \delta(r_{ij})-\frac{3}{4m_{i}m_{j}r^3_{ij}}S_{ij}\right], \\
 V^{\sigma } (r_{ij} )&=&-\frac{g_{ch}^{2} }{4\pi } \frac{\Lambda _{\sigma }^{2}m_{\sigma  }  }
 {\Lambda _{\sigma}^{2}-m_{\sigma }^{2} }\left [ Y\left ( m_{\sigma }r_{ij} \right )
  -\frac{\Lambda _{\sigma}}{m_{\sigma }}Y\left ( \Lambda _{\sigma }r_{ij} \right )\right ] \\
V^{OBE}(r_{ij}) & = & v^{\pi}(r_{ij}) \sum_{a=1}^3\boldsymbol{\lambda}_{i}^{a}\cdot
\boldsymbol{\lambda}_{j}^{a}+v^{K}(r_{ij})\sum_{a=4}^7\boldsymbol{\lambda}_{i}^{a}\cdot
\boldsymbol{\lambda} _{j}^{a} +v^{\eta}(r_{ij})\left[\left(\boldsymbol{\lambda} _{i}^{8}\cdot
\boldsymbol{\lambda} _{j}^{8}\right)\cos\theta_P-(\boldsymbol{\lambda} _{i}^{0}\cdot
\boldsymbol{\lambda}_{j}^{0}) \sin\theta_P\right] \\
&& +v^{D}(r_{ij}) \sum_{a=9}^{12}
\boldsymbol{\lambda}_{i}^{a}\cdot \boldsymbol{\lambda}_{j}^{a}+v^{D_{s}}(r_{ij}) \sum_{a=13}^{14}
\boldsymbol{\lambda}_{i}^{a}\cdot \boldsymbol{\lambda}_{j}^{a}
+v^{\eta_{c}}(r_{ij})
\boldsymbol{\lambda}_{i}^{15}\cdot \boldsymbol{\lambda}_{j}^{15}    \\
v^{\chi}(r_{ij})&=& -\frac{g_{ch}^{2} }{4\pi }\frac{m_{\chi }^{2} }{12m_{i}m_{j}}\frac{\Lambda^2}{\Lambda^2-m_{\chi}^2}m_\chi \left\{ \left[Y(m_\chi r_{ij})-\frac{\Lambda^3}{m_{\chi}^3}Y(\Lambda r_{ij})
 \right] \boldsymbol{\sigma}_i \cdot \boldsymbol{\sigma}_j  \right. \nonumber \\
&& \left.+ \left[ H(m_\chi r_{ij})-\frac{\Lambda^3}{m_\chi^3}
 H(\Lambda r_{ij}) \right] S_{ij} \right\} \boldsymbol{\lambda}^F_i \cdot
 \boldsymbol{\lambda}^F_j, ~~~\chi=\pi,K,\eta,D,D_{s},\eta_{c}\\
 S_{ij} & = &  \frac{{\boldsymbol (\sigma}_i \cdot {\boldsymbol r}_{ij})
({\boldsymbol \sigma}_j \cdot {\boldsymbol
 r}_{ij})}{r_{ij}^2}-\frac{1}{3}~{\boldsymbol \sigma}_i \cdot {\boldsymbol
\sigma}_j.
\label{H}
\end{eqnarray}
\end{widetext}
 Where $T_c$ is the kinetic energy of the center of mass; $S_{ij}$ is quark tensor operator. We only consider
 the $S-$wave systems at present, so the tensor force dose not work here; $Y(x)$ and $H(x)$ are
 standard Yukawa functions~\cite{ChQM(RPP)}; $\alpha_{ch}$ is the chiral coupling constant, determined as usual from the $\pi$-nucleon coupling constant; $\alpha_{s}$ is the quark-gluon coupling constant~\cite{ChQM1}. Here $m_{\chi}$ is the mass of the mesons, which are experimental value; $\Lambda_{\chi}$ is the cut-off parameters of different mesons, which can refer to Ref~\cite{D}. The coupling constant $g_{ch}$ for scalar chiral field is determined from the $NN\pi$ coupling constant through
 \begin{equation}
 \frac{g_{ch}^{2} }{4\pi } =(\frac{3}{5} )^{2} \frac{g_{\pi NN}^{2} }{4\pi}\frac{m_{u,d}^{2} }{m_{N}^{2} }
 \end{equation}
All other symbols have their usual meanings.
\begin{table}[ht]
\renewcommand{\arraystretch}{1.5}
\begin{center}
\caption{Model parameters} 

\begin{tabular}{lccccc} \hline \hline
& ~~~~~~$b$~~~~ & ~~~~$m_{u,d}$~~~~ & ~~~~$m_{s}$~~~~ & ~~~~$m_{c}$~~~~ & ~~~~$m_{b}$~~~~   \\
& (fm) & (MeV) & (MeV) & (MeV) & (MeV)   \\ \hline\noalign{\smallskip}
ChQM & 0.52088  & 313  &  590 & 1700 &    5105   \\ \hline\noalign{\smallskip}
 & $ a_c$ & $V_{0}$ &  $\alpha _{s_{qq} }  $ & $\alpha _{s_{qs} }  $ &  $\alpha _{s_{ss} } $   \\
 & (MeV\,fm$^{-2}$) & (fm$^{2}$) &     \\\hline\noalign{\smallskip}
ChQM & 49.350 & -1.0783  & 0.67321&   0.85644 &  0.71477  \\
\hline
 &  $\alpha _{s_{qc} }  $ & $\alpha _{s_{sc} }  $ &  $\alpha _{s_{cc} } $   \\ \hline
ChQM &0.59301  &0.60775&   1.0807\\\hline\hline
\end{tabular}
\label{parameters}
\end{center}
\end{table}

\begin{table}[ht]
\renewcommand{\arraystretch}{1.5}
\begin{center}
\caption{The calculated masses (in MeV) of the  baryons in ChQM.
 Experimental values are taken from the Particle Data Group (PDG)~\cite{PDG}. }
\begin{tabular}{lcccccc} \hline \hline
               & ~~$N$~~  & ~~$\Delta$~~  & ~~$\Lambda$~~   & ~~$\Sigma$~~ & ~~$\Sigma^{*}$~~   & ~~$\Omega$~~  \\ \hline
ChQM          & 933   & 1254     & 1100  & 1201  & 1370  & 1664     \\
 Exp.          & 939  &1233      &1116    &1189   &1315    &1672   \\ \hline
                & ~~$\Xi$~~ &~~$\Xi^{*}$~~ &  ~~$\Lambda_{c}$~~  & ~~$\Sigma_{c}$~~ &   ~~$\Sigma^{*}_{c}$~~   & ~~$\Xi_{c}$~~ \\ \hline
 ChQM           & 1338  & 1507   &2225 &  2416 &  2449       & 2450 \\
 Exp.          &1385    &1530    &2286  & 2455  &  2520      & 2470 \\ \hline
                & ~~$\Xi^{'}_{c}$~~ & ~~$\Xi^{*}_{c}$~~&~~$\Xi_{cc}$~~   &~~$\Omega_{c}$~~  &~~$\Omega^{*}_{c}$~~   &\\ \hline
ChQM           & 2546  & 2571       & 3493    &2696  &2714   \\
 Exp.           & 2578       & 2645  & 3519     &2695   &2700     \\
\hline\hline
\end{tabular}
\label{baryons}
\end{center}
\end{table}

All parameters were determined by fitting the masses of the baryons of light and heavy flavors. The model parameters and the fitting masses of baryons are shown in Table~\ref{parameters} and Table~\ref{baryons}, respectively.

\subsection{Calculation methods}
In this work, RGM~\cite{RGM1,RGM2} is used to carry out a dynamical calculation. In the framework of RGM, which split the dibaryon system into two clusters,
the main feature of RGM is that for a system consisting of two clusters, it can assume that the two clusters are frozen inside, and only consider the relative motion between the two clusters, so the conventional ansatz for the two-cluster wave function is:
\begin{equation}
\psi_{6q} = {\cal A }\left[[\phi_{B_{1}}\phi_{B_{2}}]^{[\sigma]IS}\otimes\chi_{L}(\boldsymbol{R})\right]^{J}, \label{6q}
\end{equation}
where the symbol ${\cal A }$ is the anti-symmetrization operator. With the $SU(4)$ extension, both the light and heavy quarks are considered as identical particles. So ${\cal A } = 1-9P_{36}$. $[\sigma]=[222]$ gives the total color symmetry and all other symbols have their usual meanings. $\phi_{B_{i}}$ is the $3-$quark cluster wave function. From the variational principle, after variation with respect to the relative motion wave function $\chi(\boldsymbol{\mathbf{R}})=\sum_{L}\chi_{L}(\boldsymbol{\mathbf{R}})$, one obtains the RGM equation
\begin{equation}
\int H(\boldsymbol{\mathbf{R}},\boldsymbol{\mathbf{R'}})\chi(\boldsymbol{\mathbf{R'}})d\boldsymbol{\mathbf{R'}}=E\int N(\boldsymbol{\mathbf{R}},\boldsymbol{\mathbf{R'}})\chi(\boldsymbol{\mathbf{R'}})d\boldsymbol{\mathbf{R'}}
\label{RGM}
\end{equation}
where $H(\boldsymbol{\mathbf{R}},\boldsymbol{\mathbf{R'}})$ and  $N(\boldsymbol{\mathbf{R}},\boldsymbol{\mathbf{R'}})$ are Hamiltonian and norm kernels.
The RGM can be written as
\begin{equation}
\int L(\boldsymbol{\mathbf{R}},\boldsymbol{\mathbf{R'}})\chi(\boldsymbol{\mathbf{R'}})d\boldsymbol{\mathbf{R'}}=0
\end{equation}
where
\begin{eqnarray}
 L(\boldsymbol{\mathbf{R}},\boldsymbol{\mathbf{R'}})&=& H(\boldsymbol{\mathbf{R}},\boldsymbol{\mathbf{R'}})
-EN(\boldsymbol{\mathbf{R}},\boldsymbol{\mathbf{R'}}) \nonumber \\
&=& \left [ -\frac{\bigtriangledown _{\boldsymbol{\mathbf{R'}}}^{2} }{2\mu }
+V_{rel}^{D}(\boldsymbol{\mathbf{R'}})-E_{rel}\right]\delta (\boldsymbol{\mathbf{R}}-\boldsymbol{\mathbf{R'}}) \nonumber \\
&+&H^{EX}(\boldsymbol{\mathbf{R}},\boldsymbol{\mathbf{R'}})
-EN^{EX}(\boldsymbol{\mathbf{R}},\boldsymbol{\mathbf{R'}})
\end{eqnarray}
where $\mu$ is the approximate mass between the two quark clusters; $E_{rel}=E-E_{int}$ is the relative motion energy; $V_{rel}^{D}$ is the direct term in the interaction potential.
By solving the RGM equation, we can get the energies $E$ and the wave functions. In fact, it is not convenient to work with the RGM expressions. Then, we expand the relative motion wave function $\chi(\boldsymbol{\mathbf{R}})$ by using a set of gaussians with different centers,
\begin{eqnarray}
& & \chi_{L}(\boldsymbol{R}) = \frac{1}{\sqrt{4\pi}}(\frac{3}{2\pi b^2})^{3/4} \sum_{i=1}^{n} C_{i}  \nonumber \\
&& ~~~~\times  \int \exp\left[-\frac{3}{4b^2}(\boldsymbol{R}-\boldsymbol{S}_{i})^{2}\right] Y_{LM}(\hat{\boldsymbol{S}_{i}})d\hat{\boldsymbol{S}_{i}},~~~~~
\end{eqnarray}
where $L$ is the orbital angular momentum between two clusters. Since the system we studied are all $S-$waves, $L=0$ in this work, and $\boldsymbol {S_{i}}$, $i=1,2,...,n$ are the generator coordinates, which are introduced to expand the relative motion wave function. By including the center of mass motion:
\begin{equation}
\phi_{C} (\boldsymbol{R}_{C}) = (\frac{6}{\pi b^{2}})^{3/4}e^{-\frac{3\boldsymbol{R}^{2}_{C}}{b^{2}}},
\end{equation}
the ansatz Eq.(\ref{6q}) can be rewritten as
\begin{eqnarray}
& &\psi_{6q} = {\cal A} \sum_{i=1}^{n} C_{i} \int \frac{d\hat{\boldsymbol{S}_{i}}}{\sqrt{4\pi}}
\prod_{\alpha=1}^{3}\phi_{\alpha}(\boldsymbol{S}_{i}) \prod_{\beta=4}^{6}\phi_{\beta}(-\boldsymbol{S}_{i}) \nonumber \\
& & ~~~~\times   \left[[\chi_{I_{1}S_{1}}(B_{1})\chi_{I_{2}S_{2}}(B_{2})]^{IS}Y_{LM}(\hat{\boldsymbol{S}_{i}})\right]^{J} \nonumber \\
& & ~~~~\times  [\chi_{c}(B_{1})\chi_{c}(B_{2})]^{[\sigma]}, \label{6q2}
\end{eqnarray}
where $\chi_{I_{1}S_{1}}$ and $\chi_{I_{2}S_{2}}$ are the product of the flavor and spin wave functions, and $\chi_{c}$ is the color wave function. The flavor, spin, and color wave functions are constructed in two steps. First, constructing the wave functions for the baryon and baryon clusters; then, coupling the two wave functions of two clusters to form the wave function for the dibaryon system. The detail of constructing the wave functions are presented in Appendix. For the orbital wave functions,   $\phi_{\alpha}(\boldsymbol{S}_{i})$ and $\phi_{\beta}(-\boldsymbol{S}_{i})$ are the single-particle orbital wave functions with different
reference centers:
\begin{eqnarray}
\phi_\alpha(\boldsymbol {S_{i}})=\left(\frac{1}{\pi
b^2}\right)^{\frac{3}{4}}e^ {-\frac{(\boldsymbol {r}_{\alpha}-\boldsymbol
{S_i}/2)^2}{2b^2}},
 \nonumber\\
\phi_\beta(-\boldsymbol {S_{i}})=\left(\frac{1}{\pi
b^2}\right)^{\frac{3}{4}}e^ {-\frac{(\boldsymbol {r}_{\beta}+\boldsymbol
{S_i}/2)^2}{2b^2}} .
\end{eqnarray}
By expanding the relative motion wave function between two clusters in the RGM equation by gaussians, the integro-differential equation of RGM can be reduced to an algebraic equation, which is the generalized eigen-equation. 
With the reformulated ansatz, the RGM equation Eq.(\ref{RGM}) becomes an algebraic eigenvalue equation:
\begin{equation}
\sum_{j} C_{j}H_{i,j}= E \sum_{j} C_{j}N_{i,j}.
\end{equation}
where $H_{i,j}$ and $N_{i,j}$ are the Hamiltonian matrix elements and overlaps, respectively. Besides, to keep the matrix dimension manageably small, the baryon-baryon separation is taken to be less than 6 fm in the calculation. By solving the generalized energy problem, we can obtain the energy and the corresponding wave functions of the dibaryon system. On the basis of RGM, we can further calculate scattering problems to find resonance states.

For a scattering problem, the relative wave function of the baryon-baryon is expanded as
\begin{equation}
\chi_{L} \left( \boldsymbol{R} \right) =\sum_{i=1}^{n}C_{i}\frac{\tilde{u}_{L}\left(\boldsymbol{R},\boldsymbol{S}_{i} \right)}{\boldsymbol{R}}Y_{L,M}\left ( \hat{\boldsymbol{R}}\right )
\end{equation}
with
\begin{equation}
\tilde{u}_{L}\left ( \boldsymbol{R},\boldsymbol{S}_{i}  \right )   \nonumber \\
\end{equation}
\begin{equation}
=\left\{\begin{matrix}
& \alpha _{i}u_{L}\left ( \boldsymbol{R},\boldsymbol{S}_{i} \right ), ~~~~~~~~~~~~~~~~~~~~~~&\boldsymbol{R}\le \boldsymbol{R}_{C}  \\
& \left [ h_{L}^{-}\left ( \boldsymbol{k},\boldsymbol{R}\right )-s_{i}h_{L}^{+}\left ( \boldsymbol{k},\boldsymbol{R}\right )\right ]\boldsymbol{R}, &\boldsymbol{R}\ge \boldsymbol{R}_{C}
\end{matrix}\right.
\end{equation}
where
\begin{equation}
u_{L}\left ( \boldsymbol{R} \right )=\sqrt{4\pi}\left ( \frac{3}{2\pi b^{2} }  \right )e^{-\frac{3}{4b^{2} }\left ( \boldsymbol{R}^{2}+r_{i}^{2}\right )}j_{L}\left ( -i\frac{3}{2b^{2} }Rr_{i}\right )
\end{equation}
$C_{i}$ are the expansion coefficients, and $C_{i}$ satisfy $\sum_{i=1}^{n}C_{i}=1$. n is the number of Gaussion bases (which is determined by the stability of the results), and $j_{L}$ is the $L$th spherical Bessel function. $h_{L}^{\pm}$ are the $L$th spherical Hankel functions, $k$ is the momentum of the relative motion with $k =\sqrt{2\mu E_{cm} }$, $\mu$ is the reduced mass of two baryons of the open channel, $E_{cm}$ is the incident energy of the relevant open channels, and $R_{C}$ is a cutoff radius beyond which all of the strong interactions can be disregarded. $\alpha_{i}$ and $s_{i}$ are complex parameters that determined in terms of continuity conditions at $R=R_{C}$. After performing the variational procedure by the Kohn-Hulth$\acute{e}$n-Kato(KHK) variational method~\cite{KHK}, a $L$th partial-wave equation for the scattering problem can be reduced as
\begin{equation}
\sum_{j}^{n} \mathcal{L}_{ij}^{L} C_{j} =\mathcal{M}_{ij}^{L}~~~~~~~~(i=0,1,...,n-1),
\label{M}
\end{equation}
with
\begin{equation}
\mathcal{L}_{ij}^{L}=\mathcal{K}_{ij}^{L}-\mathcal{K}_{i0}^{L}-\mathcal{K}_{0j}^{L}+\mathcal{K}_{00}^{L}
\end{equation}
\begin{equation}
\mathcal{M}_{i}^{L}=\mathcal{M}_{ij}^{L}-\mathcal{K}_{i0}^{L}
\end{equation}
and
\begin{equation}
\mathcal{K}_{ij}^{L}=\left\langle \hat{\phi_{A}}\hat{\phi_{B}}\frac{\tilde{u}_{L}(\boldsymbol{R}',\boldsymbol{S}_{i} )}{\boldsymbol{R}'}Y_{L,M}(\boldsymbol{R}')\left| H-E \right| \right.\nonumber \\
\end{equation}
\begin{equation}
\left.\cdot \mathcal{A}\left [ \hat{\phi_{A}}\hat{\phi_{B}}\frac{\tilde{u}_{L}(\boldsymbol{R},\boldsymbol{S}_{j} )}{\boldsymbol{R}}Y_{L,M}(\boldsymbol{R}) \right ] \right \rangle
\end{equation}
By solving Eq.(\ref{M}) we obtain the expansion coefficients $C_{i}$. Then, the $S$ matrix element $S_{L}$ and the phase shifts $\delta_{L}$ are given by
\begin{equation}
S_{L}\equiv e^{2i\delta_{L}}=\sum_{i=1}^{n}C_{i}S_{i}
\end{equation}

\begin{figure}
\centering
\includegraphics[height=10cm,width=15cm]{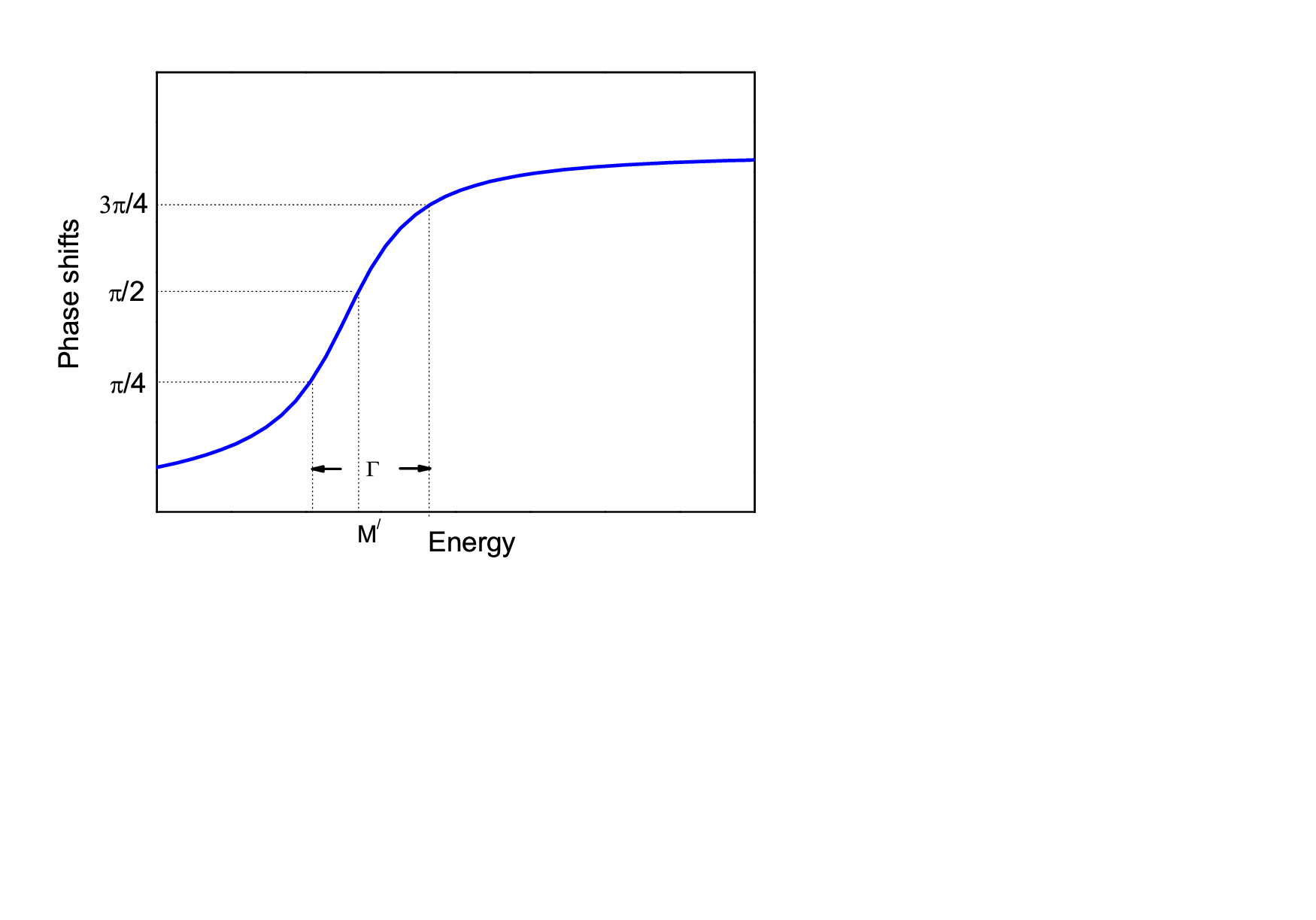}\\
\vspace{-3.5cm}
\caption{The resonance phenomena in scattering phase shifts.} \label{phaseshift}
\end{figure}

Through the scattering process, not only can we better study the interaction between hadrons, but it can also help us research resonance states. The general scattering phase shift diagram should be a smooth curve, that is, the phase shift will change gently as the incident energy increases. But in some cases, the phase shift will be abrupt, the change will be more than 90 degrees, which is the resonance phenomena. The rapid phase change is a general feature of resonance phenomena, see Fig.\ref{phaseshift}. The center of mass energy with phase shift $\frac{\pi }{2}$ gives the mass of the resonance ($M^{\prime}$ in Fig.\ref{phaseshift}), and the difference of the energies with phase shift $\frac{3\pi }{4}$ and $\frac{\pi }{4}$ gives the partial decay width of the resonance ($\Gamma$ in Fig.\ref{phaseshift}).


\section{The results and discussions}
In this work, we perform a systematical investigation of the S-wave singly charmed dibaryon systems with strange $S=-1,-3,-5$, isospin $I=0$, and the angular momentum $J=1$. To study the interaction between two hadrons, we calculate the effective potential of the system. Then, a dynamic calculation are carried out to search for bound states. Besides, the scattering process is also investigated to look for the existence of any resonance states.

\subsection{Effective potentials}
\begin{figure}
\centering
\includegraphics[height=10cm,width=15cm]{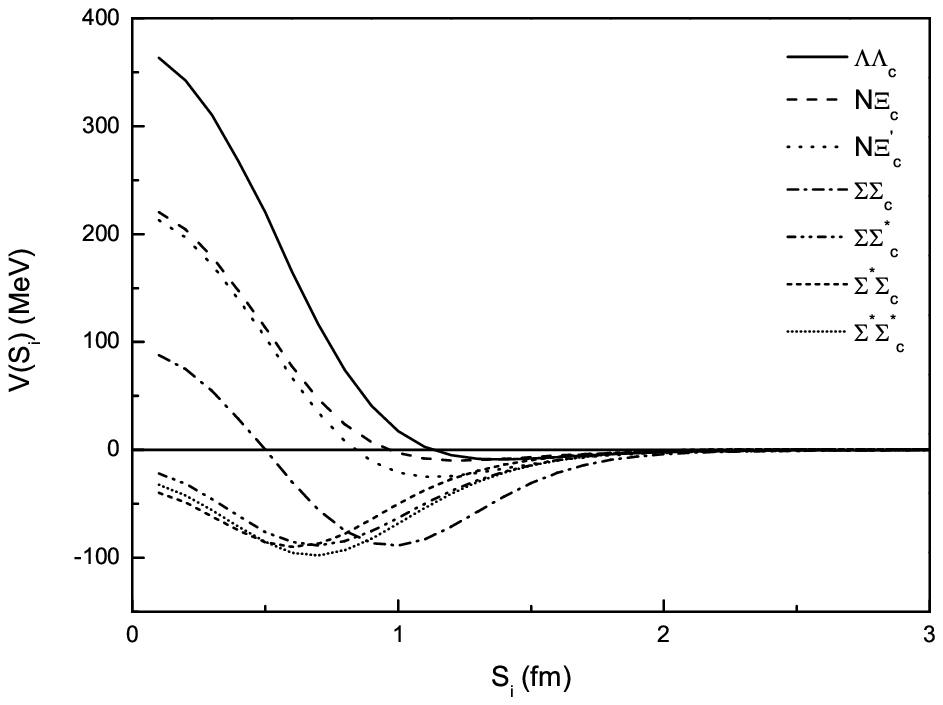}\\
\vspace{-4.0cm}
\caption{The effective potentials of different channels of the singly charmed dibaryon with $S=-1$.} \label{C1S1}
\end{figure}

\begin{figure}
\begin{center}
\includegraphics[height=9cm,width=13.5cm]{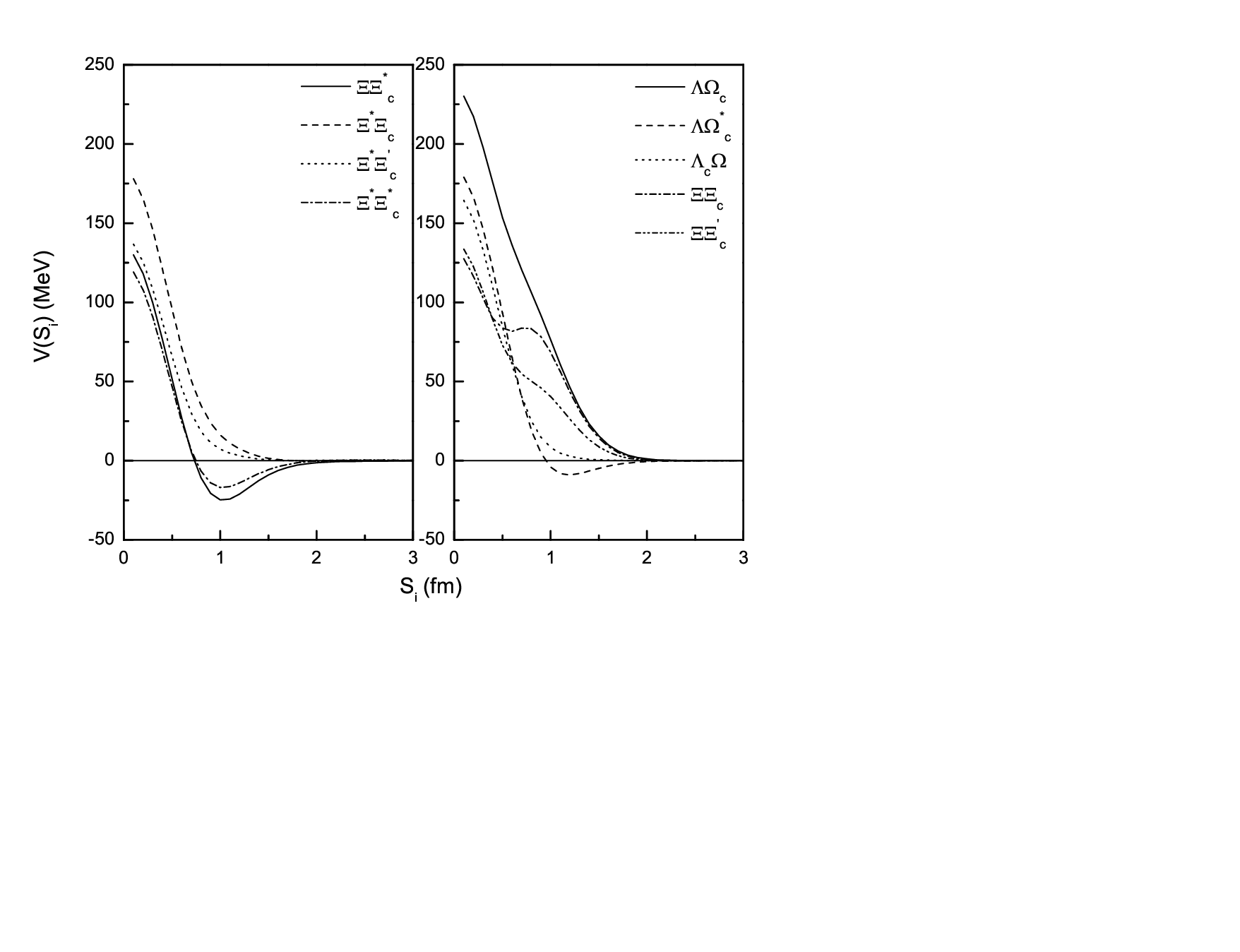}\\
\vspace{-3.5cm}
\caption{The effective potentials of different channels of the singly charmed dibaryon with $S=-3$.} \label{C1S3}
\end{center}
\end{figure}

\begin{figure}
\centering
\includegraphics[height=10cm,width=15cm]{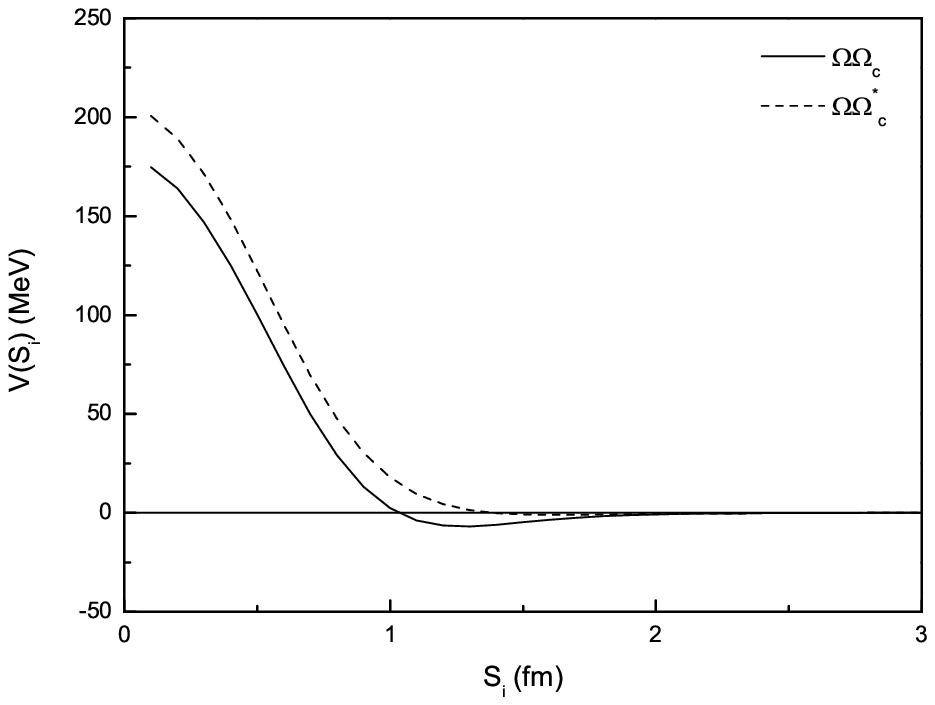} \\
\hspace{-1.0cm}
\vspace{-4.0cm}
\caption{The effective potentials of different channels of the singly charmed dibaryon with $S=-5$.} \label{C1S5}
\end{figure}

The effective potential between two baryons is shown as
\begin{equation}
V({S_{i}})=E({S_{i}})-E({\infty})
\end{equation}
where $S_{i}$ stands for the distance between two clusters and $E(\infty)$ stands for a sufficient large distance of two clusters, and the expression of $E(S_{i})$ is as follow.
\begin{equation}
E(S_{i})=\frac{\left \langle \Psi _{6q}(S_{i})\left | H \right | \Psi _{6q}(S_{i}) \right \rangle }{\left \langle \Psi _{6q}(S_{i})|\Psi _{6q}(S_{i})  \right \rangle  }
\end{equation}
$\Psi _{6q}(S_{i})$ represents the wave function of a certain channel. Besides, $ \left \langle \Psi _{6q}(S_{i})\left | H \right | \Psi _{6q}(S_{i}) \right \rangle $ and $ \left \langle \Psi _{6q}(S_{i})|\Psi _{6q}(S_{i})  \right \rangle $ are the Hamiltonian matrix and the overlap of the states.
The effective potentials of all channels with different strange numbers are shown in Fig.\ref{C1S1}, Fig.\ref{C1S3} and Fig.\ref{C1S5} respectively.

For $S=-1$ system, as shown in Fig.\ref{C1S1}, all of the seven channels are attractive, the potentials for the four channels $ \Sigma\Sigma_{c},\Sigma\Sigma_{c}^{*},\Sigma^{*}\Sigma_{c}$ and $\Sigma^{*}\Sigma_{c}^{*}$ are deeper than the other three channels $\Lambda\Lambda_{c},N\Xi_{c}$ and $N\Xi_{c}^{\prime}$, which indicates that the $\Sigma\Sigma_{c},\Sigma\Sigma_{c}^{*},\Sigma^{*}\Sigma_{c}$ and $\Sigma^{*}\Sigma_{c}^{*}$ are more likely to form bound states or resonance states.

For $S=-3$ system, from Fig.\ref{C1S3} we can see that the potentials of the $\Xi\Xi_{c}^{*},\Xi^{*}\Xi_{c}^{*}$ and $\Lambda\Omega_{c}^{*}$ are attractive, while the potentials for the other six channels are repulsive. The attraction of $\Xi\Xi_{c}^{*}$ and $\Xi^{*}\Xi_{c}^{*}$ is much stronger than that of $\Lambda\Omega_{c}^{*}$, which implies that it is more possible for $\Xi\Xi_{c}^{*}$ and $\Xi^{*}\Xi_{c}^{*}$ to form bound states or resonance states. However, compared to $S=-1$, the attraction is much weaker.

For $S=-5$ system, see Fig.\ref{C1S5}, there are only two channels in this system, one of which is $\Omega\Omega_{c}$, a purely repulsive state; and the other is $\Omega\Omega_{c}^{*}$, which is weakly attractive. Therefore, it is difficult for these channels to form any bound state. However, we still need to confirm the existence of bound states or resonance states by performing the dynamic calculations.

\subsection{Bound state calculation}
In order to see whether there is any bound state, a dynamic calculation based on RGM~\cite{RGM2} has been performed. The energies of each channel as well as the one with channel coupling calculation are listed in Table~\ref{result1}, Table~\ref{result2} and Table~\ref{result3}. The first column is the state of every channel; the second column $E_{th}$ denotes the theoretical threshold of each corresponding state; the third column $E_{sc}$ represents the energy of every single channel; the fourth column $B_{sc}$ stands for the binding energy of every single channel, which is $B_{sc}=E_{sc}-E_{th}$; the fifth column $E_{cc}$ denotes the lowest energy of the system by channel coupling calculation; and the last column $B_{cc}$ represents the binding energy with all channels coupling, which is $B_{cc}= E_{cc}-E_{th}$. Here, we should notice that when the state is unbound, we label it as ``ub". \\

\begin{table*}[ht]
\renewcommand{\arraystretch}{1.5}
\caption{The energy (in MeV) of C=1, S=-1 for the charmed dibaryon systems.}   \label{result1}
\centering
\begin{tabular}{ccccccc}
\hline \hline
   ~~ $Channels$ &  ~~$E_{th}$~(MeV) &  ~~$E_{sc}$~(MeV) &  ~~$B_{sc}$~(MeV) &    ~~$E_{cc}$~(MeV) &  ~~$B_{cc}$~(MeV)
  \\ \hline
    ~~$\Lambda\Lambda_{c}$ &    ~~3324.88 &    ~~3331.25&   ~~ub &   ~~\multirow{7}{*}{3331.21} &  ~~\multirow{7}{*}{ub}    \\ \cline{1-4}
    ~~$N\Xi_{c}$ &      ~~3383.51 &   ~~3389.46&    ~~ub &     ~~ &      \\ \cline{1-4}
   ~~${N\Xi^{'}_{c}}$ &   ~~3479.16 &   ~~3483.34 &    ~~ub  &     ~~ &      \\ \cline{1-4}
   ~~$\Sigma\Sigma_{c}$ &     ~~3617.83 &   ~~3595.26 &    ~~-22.57 &   ~~ &      \\\cline{1-4}
   ~~$\Sigma\Sigma^{*}_{c}$ &     ~~3650.64 &   ~~3625.19 &    ~~-25.45 &   ~~ &      \\\cline{1-4}
   ~~$\Sigma^{*}\Sigma_{c}$ &     ~~3786.60 &   ~~3754.77 &    ~~-31.83 &   ~~ &      \\\cline{1-4}
   ~~$\Sigma^{*}\Sigma^{*}_{c}$ &     ~~3819.41&   ~~3794.34 &    ~~-25.02 &  ~~ &      \\\cline{1-4}\hline\hline
\end{tabular}
\end{table*}
\begin{table*}[ht]
\renewcommand{\arraystretch}{1.5}
\caption{The energy (in MeV) of C=1, S=-3 for the charmed dibaryon systems.}   \label{result2}
\centering
\begin{tabular}{ccccccc}
\hline \hline
   ~~ $Channels$ &  ~~$E_{th}$~(MeV) &  ~~$E_{sc}$~(MeV) &  ~~$B_{sc}$~(MeV) &    ~~$E_{cc}$~(MeV) &  ~~$B_{cc}$~(MeV)
  \\ \hline
  ~~${\Lambda\Omega_{c}}$ &    ~~3796.20&     ~~3802.83 &   ~~ub &   ~~\multirow{9}{*}{3789.92} &   ~~\multirow{9}{*}{ub}   \\ \cline{1-4}
    ~~$\Lambda\Omega^{*}_{c}$ &   ~~3813.99 &   ~~3818.86 &    ~~ub &    ~~ &     \\ \cline{1-4}
    ~~$\Lambda_{c}\Omega$ &   ~~3888.73 &   ~~3894.12 &    ~~ub &     ~~ &     \\ \cline{1-4}
~~$\Xi\Xi_{c}$ &   ~~3788.70 &   ~~3795.15 &    ~~ub  &     ~~ &     \\\cline{1-4}
~~$\Xi\Xi^{'}_{c}$ &   ~~3884.30 &   ~~3890.53 &    ~~ub  &     ~~ &     \\\cline{1-4}
~~$\Xi\Xi^{*}_{c}$ &   ~~3909.60 &   ~~3912.23 &    ~~ub  &     ~~ &     \\\cline{1-4}
~~$\Xi^{*}\Xi_{c}$ &   ~~3957.42 &   ~~3963.12 &    ~~ub  &     ~~ &     \\\cline{1-4}
~~$\Xi^{*}\Xi^{'}_{c}$ &   ~~4053.07 &   ~~4058.67 &    ~~ub  &     ~~ &     \\\cline{1-4}
~~$\Xi^{*}\Xi^{*}_{c}$ &   ~~4078.37&   ~~4082.40 &    ~~ub  &    ~~ &     \\\cline{1-4}
 \hline\hline
\end{tabular}
\end{table*}
\begin{table*}[ht]
\renewcommand{\arraystretch}{1.5}
\caption{The energy (in MeV) of C=1, S=-5 for the charmed dibaryon systems.}   \label{result3}
\centering
\begin{tabular}{ccccccc}
\hline \hline
   ~~ $Channels$ &  ~~$E_{th}$~(MeV) &  ~~$E_{sc}$~(MeV) &  ~~$B_{sc}$~(MeV) &    ~~$E_{cc}$~(MeV) &  ~~$B_{cc}$~(MeV)
  \\ \hline
 ~~${\Omega\Omega_{c}}$ &  ~~4360.05 &    ~~4363.77 &   ~~ub &    ~~\multirow{2}{*}{4363.77} &   ~~\multirow{2}{*}{ub} \\ \cline{1-4}
~~${\Omega\Omega^{*}_{c}}$ &   ~~4377.84&   ~~4382.12 &    ~~ub  &    ~~ &     \\\cline{1-4}
 \hline\hline
\end{tabular}
\end{table*}

 \textbf{\textit{S=-1}}:
  The single channel calculation shows that the channels $\Sigma\Sigma_{c}, \Sigma\Sigma_{c}^{*}, \Sigma^{*}\Sigma_{c}$ and $\Sigma^{*}\Sigma_{c}^{*}$ are bound states with the binding energies -23 MeV, -25 MeV, -32 MeV  and -25 MeV, respectively (see Table~\ref{result1}). This conclusion is consistent with the property that there is a strong effective attraction of these channels.
  However, for the $\Lambda\Lambda_{c},N\Xi_{c}$ and $N\Xi_{c}^{\prime}$ channels, which are unbound, the energies obtained by single channel calculations are above their corresponding thresholds due to the weak attraction of these channels. For the calculation of the channel coupling, the lowest energy is still above the lowest threshold ($\Lambda\Lambda_{c}$). Therefore, for this system, no bound states below the lowest threshold were found. For higher-energy single-channel bound states, they can be coupled to the open channels and the scattering process is needed to determine the existence of resonance states.

 \textbf{\textit{S=-3}}:
 From Table~\ref{result2}, the single channel calculation shows that all these nine channels are unbound. After the channel coupling calculation, the lowest energy of this system is 3790 MeV (still higher than the threshold of the lowest channel $\Xi\Xi_{c}$ ), which indicates that the singly charmed dibaryon system with $IJ=01, S=-3$ is unbound. This is reasonable. The attractions of the channels $\Xi\Xi_{c}^{*},\Xi^{*}\Xi_{c}^{*}$ and $\Lambda\Omega_{c}^{*}$ are not strong enough to form any bound state, and the interaction of the other channels are repulsive, as shown in Fig.\ref{C1S3}.

\textbf{\textit{S=-5}}: The situation is similar to that of the $S=-3$ system. As shown in Table~\ref{result3}, both of the channels $\Omega\Omega_{c}$ and $\Omega\Omega_{c}^{*}$ are unbound. The lowest energy of the system is higher than the threshold of the $\Omega\Omega_{c}$ by the channel coupling calculation. So the system with $S=-5$ is unbound.

\subsection{Resonance states}
As mentioned above, some channels are bound due to the strong attractions of the system. However, these states will decay to the corresponding open channels by coupling with them and become resonance states. Besides, some states will become scattering state by the effect of coupling to both the open and closed channels. To further check the existence of the resonance states, we studied the scattering phase shifts of all possible open channels. Since no resonance states are obtained in the $S=-3$ and $S=-5$ systems, we only show the scattering phase shifts of the $S=-1$ system here.

In the $S=-1$ system, four singly bound states are obtained, which are $ \Sigma\Sigma_{c}$, $\Sigma\Sigma_{c}^{*}$, $\Sigma^{*}\Sigma_{c}$ and $\Sigma^{*}\Sigma_{c}^{*}$, and there are three open channels, which are $\Lambda\Lambda_{c},N\Xi_{c}$ and $N\Xi_{c}^{\prime}$. We analyze two types of channel coupling in this
work. The first is the two-channel coupling with a singly bound state and a related open channel, while the other is the five-channel coupling with four bound states and a
corresponding open channel. The general features of the calculated results are as follows.

\begin{figure}
\begin{flushleft}
\includegraphics[width=0.8\textwidth]{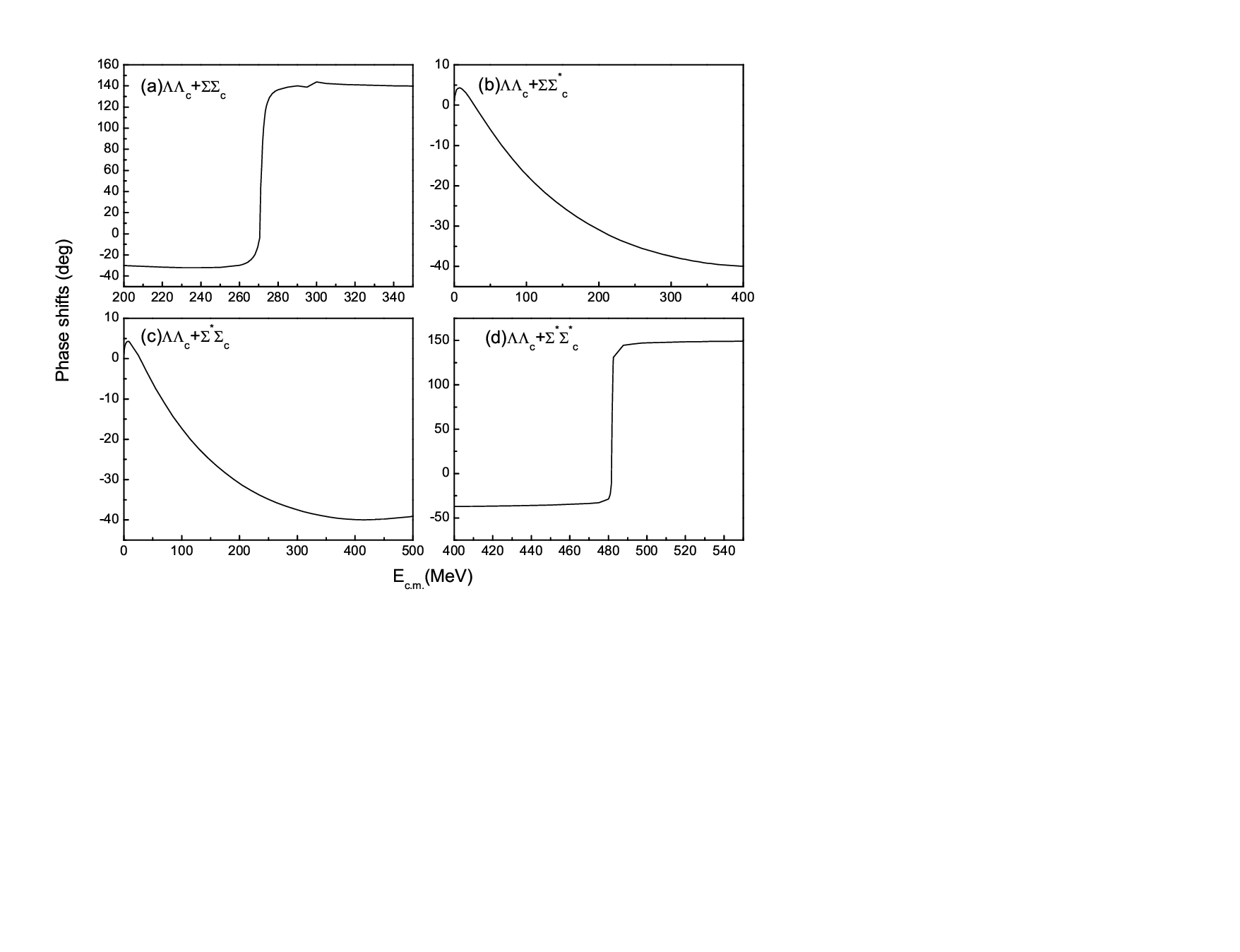}\\
\hspace{-2cm}
\vspace{-5cm}
\caption{The $\Lambda\Lambda_{c}$ phase shift with two-channel coupling for the $S=-1$ system.} \label{14567}
\end{flushleft}
\end{figure}

Here, we should note that the horizontal axis $E_{c.m.}$ in Fig.\ref{14567} is the incident energy without the theoretical threshold of the corresponding open channel. So the resonance mass $M^{\prime}$ is obtained by adding $E_{c.m.}$ and the theoretical threshold of the corresponding open channel. In order to minimize the theoretical errors and compare our predictions with future experimental data, we shift the resonance mass by $M=M^{\prime}-E_{th}+E_{exp}$, where $E_{th}$ and $E_{exp}$ are the theoretical and experimental thresholds of the resonance state, respectively. Taking the resonance state $\Lambda\Lambda_{c}$ in the $\Sigma\Sigma_{c}$ channel as  an example, the resonance mass shown in Fig.\ref{14567}(a) is $M^{\prime}=3597$ MeV, the theoretical threshold is $M_{th}=3595$ MeV, and the experimental threshold is $M_{exp}=3618$ MeV. Then the final resonance mass $M=3597-3595+3618=3620$ MeV. The estimated masses and widths of the resonances in different channels are listed in Table\ref{width}, where $M$ is the resonance mass, $\Gamma_{i}$ is the partial decay width of the resonance state decaying to different open channels, and $\Gamma_{total}$ is the total decay width of the resonance state.

For the case of the two-channel coupling, in $\Lambda\Lambda_{c}$ scattering process, it is obvious that $ \Sigma\Sigma_{c}$ and $\Sigma^{*}\Sigma_{c}^{*}$ appear as resonance states, as shown in Fig.\ref{14567}(a) and Fig.\ref{14567}(d), respectively. The resonance mass and decay width of every resonance state are obtained from the $\Lambda\Lambda_{c}$ scattering phase shifts. At the same time, $\Sigma\Sigma_{c}^{*}$ and $\Sigma^{*}\Sigma_{c}$ do not behave as resonance states in $\Lambda\Lambda_{c}$ scattering process, as shown in Fig.\ref{14567}(b) and Fig.\ref{14567}(c), respectively. There may be two reasons: the one is that stronger coupling between the two channels causes the bound state to be pushed above the threshold and become a scattering state; the other one is that the coupling between the two channels is so weak that the resonance state does not manifest during the scattering process. To clarify this issue, we calculate the cross matrix elements between the two channels ($\Lambda\Lambda_{c}$ and $\Sigma\Sigma_{c}^{*}$/$\Sigma^{*}\Sigma_{c}$), they are all close to zero, which means that the coupling between $\Lambda\Lambda_{c}$ and $\Sigma\Sigma_{c}^{*}$/$\Sigma^{*}\Sigma_{c}$ is very weak. Therefore, neither $\Sigma\Sigma_{c}^{*}$ nor $\Sigma^{*}\Sigma_{c}$ behaves as a resonance state in the $\Lambda\Lambda_{c}$ scattering phase shifts.

\begin{figure}
\begin{flushleft}
\includegraphics[width=0.8\textwidth]{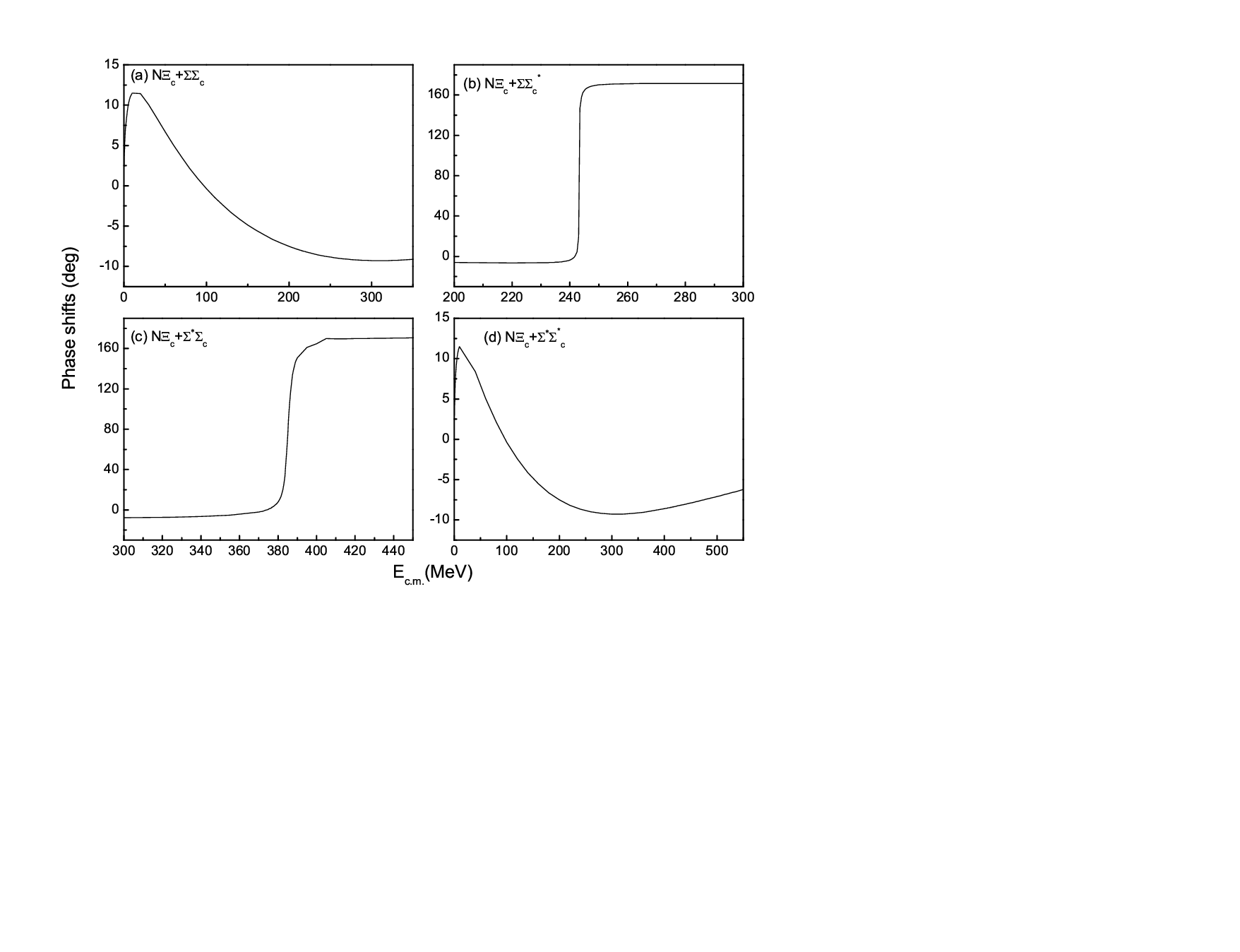}\\
\hspace{-2cm}
\vspace{-5cm}
\caption{The $N\Xi_{c}$ phase shift with two-channel coupling for the $S=-1$ system.} \label{24567}
\end{flushleft}
\end{figure}

\begin{figure}
\includegraphics[width=0.8\textwidth]{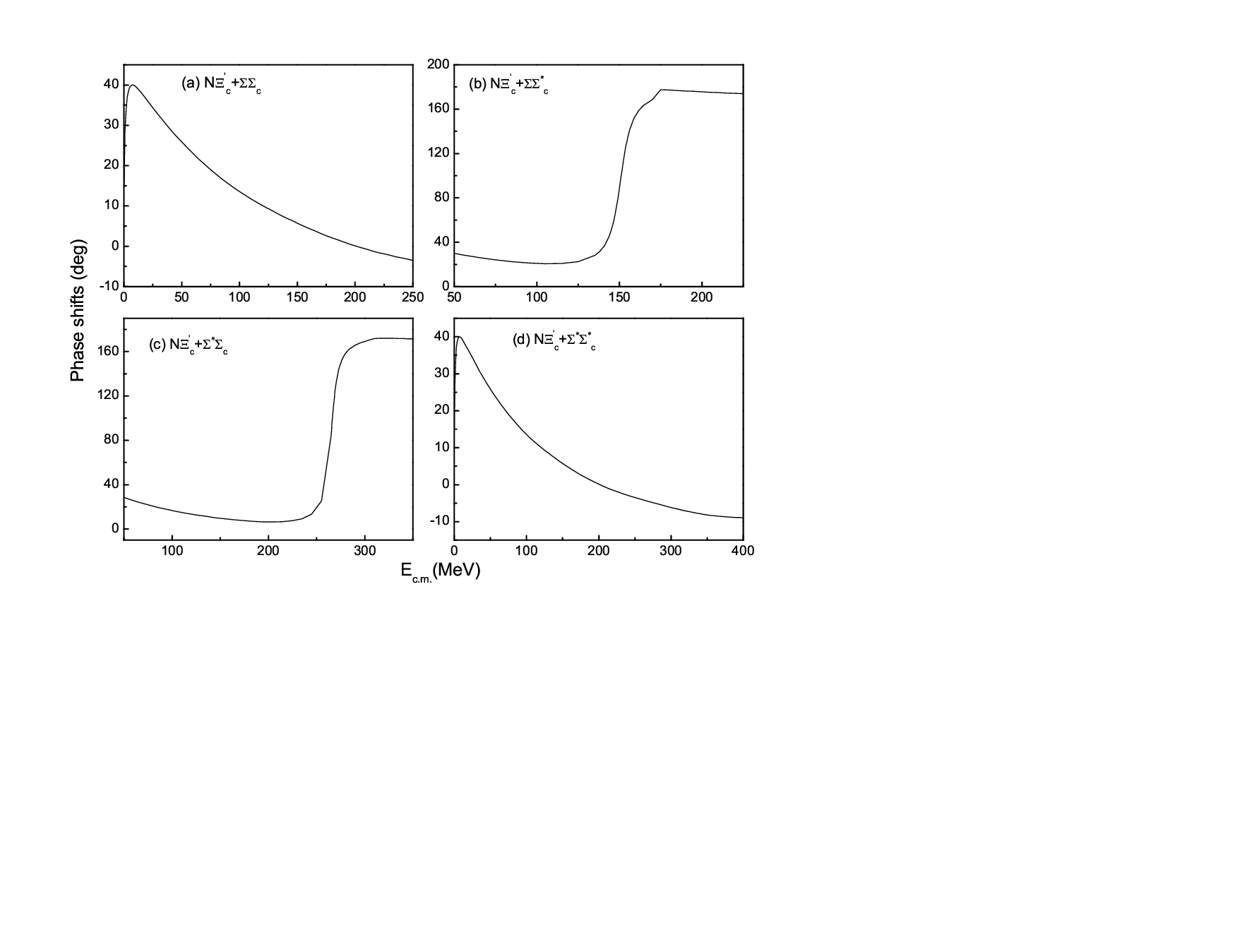}\\
\hspace{-2cm}
\vspace{-5cm}
\caption{The $N\Xi_{c}^{\prime}$ phase shift with two-channel coupling for the $S=-1$ system.} \label{34567}
\end{figure}

However, in the $N\Xi_{c}$ scattering process, the situation reversed. From Fig.\ref{24567}, both $\Sigma\Sigma_{c}^{*}$ and $\Sigma^{*}\Sigma_{c}$ appear as resonance states, while the other two channels $\Sigma\Sigma_{c}$ and $\Sigma^{*}\Sigma_{c}^{*}$ do not. The cross matrix elements between $N\Xi_{c}$ and $\Sigma\Sigma_{c}$/$\Sigma^{*}\Sigma_{c}^{*}$ show that the coupling between them is very weak, which results in the absence of resonance state $\Sigma\Sigma_{c}$/$\Sigma^{*}\Sigma_{c}^{*}$ in the $N\Xi_{c}$ scattering phase shift. In the $N\Xi_{c}^{\prime}$ scattering process, as shown in Fig.\ref{34567}, the conclusion is similar to the one in the $N\Xi_{c}$ scattering process. All the resonance masses and decay width are shown in Table\ref{width}.

\begin{figure}
\includegraphics[width=0.8\textwidth]{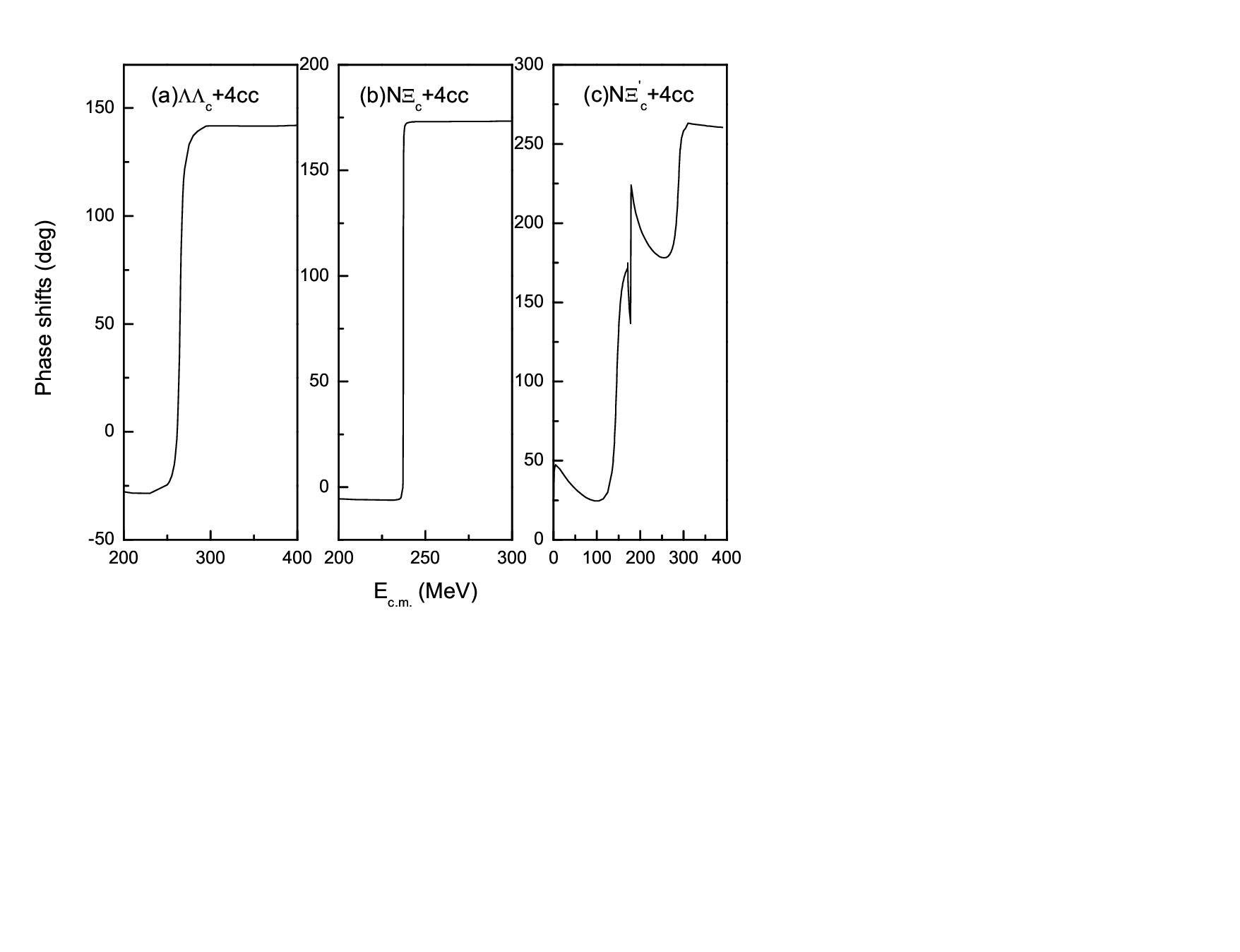}\\
\hspace{-2cm}
\vspace{-4.5cm}
\caption{The $\Lambda\Lambda_{c}$, $N\Xi_{c}$ and $N\Xi_{c}^{\prime}$ phase shifts with five-channel coupling for the $S=-1$ system.} \label{4cc}
\end{figure}

For the case of five-channel coupling, the scattering phase shifts are shown in Fig.\ref{4cc}, and the resonance masses and decay widths are listed in Table\ref{width}.
There is only one resonance state $\Sigma\Sigma_{c}$ appears in the $\Lambda\Lambda_{c}$ phase shifts, as shown in Fig.\ref{4cc}(a). From Table\ref{width}, the resonance mass of the $\Sigma\Sigma_{c}$ in the five-channel coupling case is 3591 MeV, which is lower than the one in the two-channel coupling case (3620 MeV). This is because the coupling between closed channels will push down the channels with lower energy. At the same time, the channel coupling can also raises the energy of the higher state, even pushes the higher state above the threshold. Therefore, the resonance state $\Sigma^{*}\Sigma_{c}^{*}$ in two-channel coupling disappears in the five-channel coupling. Similarly, there is only one resonance state $\Sigma\Sigma_{c}^{*}$ appears in the $N\Xi_{c}$ phase shifts, which is shown in Fig.\ref{4cc}(b). In the $N\Xi_{c}^{\prime}$ phase shifts, the situation is slightly different. Two resonance states $\Sigma\Sigma_{c}^{*}$ and $\Sigma^{*}\Sigma_{c}$ are shown in Fig.\ref{4cc}(c). By comparing with the results in the two-channel coupling, the resonance mass of $\Sigma\Sigma_{c}^{*}$ is 30 MeV lower, while the one of $\Sigma^{*}\Sigma_{c}$ is 9 MeV higher.
However, since the resonance $\Sigma^{*}\Sigma_{c}$ disappears in the $N\Xi_{c}$ scattering phase shifts, it will decay through the $N\Xi_{c}$ open channel. So the $\Sigma^{*}\Sigma_{c}$ cannot be identified as a resonance state.
All these results show that the existence of the resonance states and the resonance energy are both affected by the multi-channel coupling. So the effect of the channel coupling cannot be ignored in the multi-quark system.

\begin{table*}[htb]
\begin{center}
\renewcommand{\arraystretch}{1.5}
\caption{\label{width} The masses ($M$) and decay widths (in the unit of MeV) of resonance states with the difference scattering process. $\Gamma_{i}$ is the partial decay width of the resonance state decaying to the $i-$th open channel. $\Gamma_{total}$ is the total decay width of the resonance state.}
\begin{tabular}{p{2.0cm}<\centering p{1.cm}<\centering p{0.5cm}<\centering p{0.01cm}<\centering p{1.cm}<\centering p{0.5cm}<\centering p{0.01cm}<\centering p{1.cm}<\centering p{0.5cm}<\centering p{0.01cm}<\centering p{1.cm}<\centering p{0.5cm}<\centering p{0.01cm}<\centering p{1.cm}<\centering p{0.5cm}<\centering p{0.01cm}<\centering p{1.cm}<\centering p{0.5cm}<\centering p{0.01cm}<\centering p{1.cm}<\centering p{0.5cm}<\centering p{0.01cm}<\centering p{1.cm}<\centering p{0.5cm}<\centering p{0.01cm}<\centering p{1.cm}<\centering p{0.5cm}<\centering p{3.cm}<\centering }
\toprule[1pt]
\multirow{3}{*}{Open channels} & \multicolumn{10}{c}{Two channel coupling} & & \multicolumn{12}{c}{Five channel coupling} \\
\cline{2-12} \cline{14-24}
   &\multicolumn{2}{c}{$\Sigma\Sigma_{c}$} &  &\multicolumn{2}{c}{$\Sigma\Sigma_{c}^{\ast}$} &  &\multicolumn{2}{c}{$\Sigma^{\ast}\Sigma_{c}$} &  &\multicolumn{2}{c}{$\Sigma^{\ast} \Sigma_{c}^{\ast}$} & &\multicolumn{2}{c}{$\Sigma\Sigma_{c}$} &  &\multicolumn{2}{c}{$\Sigma\Sigma_{c}^{\ast}$} &  &\multicolumn{2}{c}{$\Sigma^{\ast}\Sigma_{c}$} &  &\multicolumn{2}{c}{$\Sigma^{\ast} \Sigma_{c}^{\ast}$} \\
   \cline{2-3}\cline{5-6}\cline{8-9}\cline{11-12}\cline{14-15}\cline{17-18}\cline{20-21}\cline{23-24}
                      & $M$  & $\Gamma_{i}$  & &$M$ & $\Gamma_{i}$ & & $M$ & $\Gamma_{i}$  &  & $M$ & $\Gamma_{i}$ & & $M$  &$\Gamma_{i}$ & &$M$ & $\Gamma_{i}$  & &$M$ & $\Gamma_{i}$ & &$M$ & $\Gamma_{i}$ \\
 \midrule[1pt]
  $\Lambda\Lambda_{c}$            &3620  &8.5  &   &$\dots$   &$\dots$ &  &$\dots$  &$\dots$  & &3832  &10.7 & &$3591$  &$11.1$ &  &$\dots$  &$\dots$  &  &$\dots$  &$\dots$  &  &$\dots$  &$\dots$   \\
  $N \Xi_{c}$     &$\dots$  &$\dots$  &   &3652   &0.6 &  &3801   &3.7  & &$\dots$   &$\dots$ & &$\dots$  &$\dots$ &  &3621  &0.1  &  &$\dots$  &$\dots$  &  &$\dots$ &$\dots$    \\
  $N \Xi_{c}^{\prime}$     &$\dots$  &$\dots$  &   &3654   &11.5 &  &3776   &10.5  & &$\dots$     &$\dots$ & &$\dots$  &$\dots$ &  &3624 &14.8  &  &3785  &32.0  &  &$\dots$  &$\dots$   \\
  $\Gamma_{total}$     &      &8.5 &   &   & 12.1 &  &  & 14.2 &  &        &  10.7     & &    &11.1 &    &     &14.9   &  &     & 32.0    &   &      &     \\
\bottomrule[1pt]
\end{tabular}
\end{center}
\end{table*}

\newpage

\section{Summary}
The S-wave singly charmed dibaryon systems with strangeness numbers $S=-1$, $-3$ and $-5$ are systemically investigated by using the RGM in the framework of ChQM. Our goal is to search for any bound state or resonance state of singly charmed dibaryon systems. Herein, the effective potentials are calculated to explore the interactions of between two baryons. Both the single-channel and the coupled-channel dynamic bound-state calculations are carried out to search for possible
states. Meanwhile, the study of the scattering process of the open channels is carried out to confirm possible resonance states.

According to the numerical results, in the $S=-1$ system, the attractions between $\Sigma$/$\Sigma^{*}$ and $\Sigma_{c}$/$\Sigma_{c}^{*}$ are large enough to form singly bound states $\Sigma\Sigma_{c}$, $\Sigma\Sigma^{*}_{c}$, $\Sigma^{*}\Sigma_{c}$ and $\Sigma^{*}\Sigma_{c}^{*}$. However, these states can couple with the corresponding open channels, and become resonance states or scattering states. By including the effect of channel-coupling, two resonance states with strangeness numbers $S=-1$ are obtained. The one is the $\Sigma\Sigma_{c}$ state with the mass and width 3591 MeV and 11.1 MeV, respectively, and the decay channel is $\Lambda\Lambda_{c}$. The other is the $\Sigma\Sigma^{\ast}_{c}$ state with the mass and width 3621-3624 MeV and 14.9 MeV, respectively, and the decay channels are $N\Xi_{c}$ and $N\Xi^{\prime}_{c}$.  All these dibaryons are worth searching for in experiments, although it will be a challenging subject.

In the past two decades, numerous heavy-flavor hadrons have been discovered in experiments, which are considered as promising candidates for tetraquarks and pentaquarks. In Ref.~\cite{Wang:2024riu}, the authors claimed that the existence of molecular states in $DD^{*}$, $D\bar{D}^{*}$, and $\Sigma_{c}\bar{D}^{(*)}$ systems leads to the emergence of a large number of deuteronlike hexaquarks in the heavy flavor sectors. The systems composed of charmed baryons and hyperons are predicted by the mass spectra calculation. In Ref.~\cite{Kong:2022rvd}, the charmed-strange molecular dibaryons are investigated in a quasipotential Bethe-Salpeter approach together with the one-boson-exchange model. The results suggested that attractions widely exist in charmed-strange system, and the S-wave bound states can be produced
from most of the channels. In this work, fewer charmed dibaryon resonance states are obtained, since the coupling with the open channels are considered.
The study of the scattering process is an effective way to look for the genuine resonances. However, to distinguish the various
explanations and confirm the existence of the exotic hadron states is still very difficult, and requires the joint
efforts of both theorists and experimentalists.

\acknowledgments{This work is supported partly by the National Natural Science Foundation of China under Contracts Nos. 11675080, 11775118 and 11535005.}

\section{Appendix}
 Here, we only list the wave functions we used in this work. The spin wave function of a $q^{3}$ cluster is labeled as $\chi_{s,s_{z}}^{\sigma}$, where $s$ and $s_{z}$ are the spin quantum number and the third component, respectively. For wave functions with the same quantum number but different symmetries, we distinguish them with different numbers. For example, $\chi_{\frac{1}{2},\frac{1}{2}}^{\sigma1}$ and $\chi_{\frac{1}{2},\frac{1}{2}}^{\sigma2}$ represent respectively the symmetric and antisymmetric spin wave functions with spin quantum number $\frac{1}{2}$.
 \begin{eqnarray}
	\chi_{\frac{3}{2}, \frac{3}{2}}^{\sigma} & = &\alpha \alpha \alpha \nonumber \\
	\chi_{\frac{3}{2}, \frac{1}{2}}^{\sigma} & = &\frac{1}{\sqrt{3}}(\alpha \alpha \beta+\alpha \beta \alpha+\beta \alpha \alpha) \nonumber \\
	\chi_{\frac{3}{2},-\frac{1}{2}}^{\sigma} & = &\frac{1}{\sqrt{3}}(\alpha \beta \beta+\beta \alpha \beta+\beta \beta \alpha) \nonumber \\
    \chi_{\frac{3}{2},-\frac{3}{2}}^{\sigma} & = &\beta \beta \beta \nonumber \\
    \chi_{\frac{1}{2},\frac{1}{2}}^{\sigma1} & = &\sqrt{\frac{1}{6}}(2 \alpha \alpha \beta-\alpha \beta \alpha-\beta \alpha \alpha)  \nonumber\\
    \chi_{\frac{1}{2},\frac{1}{2}}^{\sigma2} & = &\sqrt{\frac{1}{2}}(\alpha \beta \alpha-\beta \alpha \alpha)\nonumber \\
    \chi_{\frac{1}{2},-\frac{1}{2}}^{\sigma1} & = &\sqrt{\frac{1}{6}}(\alpha \beta \beta+\beta \alpha \beta-2 \beta \beta \alpha) \nonumber \\
    \chi_{\frac{1}{2},-\frac{1}{2}}^{\sigma2} & = &\sqrt{\frac{1}{2}}(\alpha \beta \beta-\beta \alpha \beta) \nonumber
\end{eqnarray}

The flavor wave functions of the $q^{3}$ cluster $\chi_{I, I_z}^{f}$ ($I$ and $I_z$ are the isospin quantum number and the third component, respectively) are as follows. Here, both the light and heavy quarks are considered as identical particles with the $SU(4)$ extension.
\begin{eqnarray}
    \chi_{0,0}^{f1}  & =& \frac{1}{2}(usd+sud-sdu-dsu) \nonumber \\   
    \chi_{0,0}^{f2}  & = &\sqrt{\frac{1}{12}}(2uds-2dsu+sdu+usd-sud-dsu) \nonumber \\
    \chi_{0,0}^{f3}  & = &\frac{1}{2}(ucd+cud-cdu-dcu) \nonumber \\  
    \chi_{0,0}^{f4}  & = &\sqrt{\frac{1}{12}}(2ucd-2dcu+cdu+ucd-cud-dcu) \nonumber \\
    \chi_{0,0}^{f5}  & = &\sqrt{\frac{1}{6}}(2ssc-scs-css) \nonumber \\  
    \chi_{0,0}^{f6}  & = &\sqrt{\frac{1}{2}}(scs-css) \nonumber \\
    \chi_{0,0}^{f7}  & = &\sqrt{\frac{1}{3}}(ssc+scs+css) \nonumber \\   
    \chi_{0,0}^{f8}  & = &sss          \nonumber \\   
    \chi_{\frac{1}{2},-\frac{1}{2}}^{f1}  & =&\frac{1}{2}(dcs+cds-csd-scd) \nonumber \\ 
    \chi_{\frac{1}{2},-\frac{1}{2}}^{f2}  & =&\sqrt{\frac{1}{12}}(2dsc-2sdc+csd+dcs-cds-scd) \nonumber\\
    \chi_{\frac{1}{2},-\frac{1}{2}}^{f3}  & =&\sqrt{\frac{1}{6}}(udd+dud-2ddu) \nonumber \\  
    \chi_{\frac{1}{2},-\frac{1}{2}}^{f4}  & =&\sqrt{\frac{1}{2}}(udd-dud) \nonumber \\
    \chi_{\frac{1}{2},-\frac{1}{2}}^{f5}  & =&\sqrt{\frac{1}{12}}(2dsc+2sdc-csd-dcs-cds-scd) \nonumber \\
    \chi_{\frac{1}{2},-\frac{1}{2}}^{f6}  & =&\frac{1}{2}(dcs+scd-csd-cds) \nonumber \\
    \chi_{\frac{1}{2},-\frac{1}{2}}^{f7}  & =&\sqrt{\frac{1}{6}}(dss+sds-2ssd) \nonumber \\  
    \chi_{\frac{1}{2},-\frac{1}{2}}^{f8}  & =&\sqrt{\frac{1}{2}}(dss-sds) \nonumber \\
    \chi_{\frac{1}{2},-\frac{1}{2}}^{f9}  & =&\sqrt{\frac{1}{6}}(dsc+sdc+csd+dcs+cds+scd) \nonumber\\  
    \chi_{\frac{1}{2},-\frac{1}{2}}^{f10}  & =&\sqrt{\frac{1}{3}}(dss+sds+ssd) \nonumber 
\end{eqnarray}
\begin{eqnarray}    
    \chi_{\frac{1}{2},\frac{1}{2}}^{f1}  & =&\sqrt{\frac{1}{6}}(2uud-udu-duu) \nonumber \\  
    \chi_{\frac{1}{2},\frac{1}{2}}^{f2}  & =&\sqrt{\frac{1}{2}}(udu-duu) \nonumber \\
    \chi_{\frac{1}{2},\frac{1}{2}}^{f3}  & =&\frac{1}{2}(ucs+cus-csu-scu) \nonumber \\
    \chi_{\frac{1}{2},\frac{1}{2}}^{f4}  & =&\sqrt{\frac{1}{12}}(2usc-2suc+csu+ucs-cus-scu) \nonumber\\
    \chi_{\frac{1}{2},\frac{1}{2}}^{f5}  & =&\sqrt{\frac{1}{12}}(2usc+2suc-csu-ucs-cus-scu) \nonumber\\
    \chi_{\frac{1}{2},\frac{1}{2}}^{f6}  & =&\frac{1}{2}(ucs+scu-csu-cus) \nonumber\\
        \chi_{\frac{1}{2},\frac{1}{2}}^{f7}  & =&\sqrt{\frac{1}{6}}(uss+sus-2ssu) \nonumber \\  
    \chi_{\frac{1}{2},\frac{1}{2}}^{f8}  & =&\sqrt{\frac{1}{2}}(uss-sus) \nonumber \\
    \chi_{\frac{1}{2},\frac{1}{2}}^{f9}  & =&\sqrt{\frac{1}{6}}(usc+suc+csu+ucs+cus+scu) \nonumber\\
    \chi_{\frac{1}{2},\frac{1}{2}}^{f10}  & =&\sqrt{\frac{1}{3}}(uss+sus+ssu)  \nonumber \\ 
    \chi_{1,-1}^{f1}  & = &\sqrt{\frac{1}{6}}(2ddc-dcd-cdd) \nonumber \\  
    \chi_{1,-1}^{f2}  & = &\sqrt{\frac{1}{2}}(dcd-cdd) \nonumber \\
    \chi_{1,-1}^{f3}  & = &\sqrt{\frac{1}{6}}(2dds-dsd-sdd) \nonumber \\  
    \chi_{1,-1}^{f4}  & = &\sqrt{\frac{1}{2}}(dsd-sdd) \nonumber \\
    \chi_{1,-1}^{f5}  & = &\sqrt{\frac{1}{3}}(ddc+dcd+cdd) \nonumber \\  
    \chi_{1,-1}^{f6}  & = &\sqrt{\frac{1}{3}}(dds+dsd+sdd) \nonumber\\   
    \chi_{1,0}^{f1}  & = &\sqrt{\frac{1}{12}}(2uds+2dus-sdu-usd-sud-dsu) \nonumber \\  
    \chi_{1,0}^{f2}  & = &\frac{1}{2}(usd+dsu-sdu-sud) \nonumber\\
    \chi_{1,0}^{f3}  & = &\sqrt{\frac{1}{12}}(2udc+2duc-cdu-ucd-cud-dcu) \nonumber \\  
    \chi_{1,0}^{f4}  & = &\frac{1}{2}(ucd+dcu-cdu-cud) \nonumber \\
    \chi_{1,0}^{f5}  & = &\sqrt{\frac{1}{6}}(udc+duc+cdu+ucd+cud+dcu) \nonumber \\  
    \chi_{1,0}^{f6}  & = &\sqrt{\frac{1}{6}}(uds+dus+sdu+usd+sud+dsu) \nonumber \\  
    \chi_{1,1}^{f1}  & = &\sqrt{\frac{1}{6}}(2uus-usu-suu) \nonumber \\  
    \chi_{1,1}^{f2}  & = &\sqrt{\frac{1}{2}}(usu-suu) \nonumber\\
    \chi_{1,1}^{f3}  & = &\sqrt{\frac{1}{6}}(2uuc-ucu-cuu) \nonumber   
\end{eqnarray}
\begin{eqnarray}
    \chi_{1,1}^{f4}  & = &\sqrt{\frac{1}{2}}(ucu-cuu) \nonumber \\
    \chi_{1,1}^{f5}  & = &\sqrt{\frac{1}{3}}(uuc+ucu+cuu) \nonumber \\  
    \chi_{1,1}^{f6}  & = &\sqrt{\frac{1}{3}}(uus+usu+suu) \nonumber 
\end{eqnarray}

The color wave function of a color-singlet $q^{3}$ cluster is:
\begin{align}
	\chi^{c} =& \sqrt{\frac{1}{6}}(r g b-r b g+g b r-g r b+b r g-b g r) \nonumber
\end{align}

The total flavor-spin-color wave function of the dibaryon system can be acquired by substituting the wave functions of the flavor, the spin, and the color parts according to the given quantum number of the system, and the total flavor-spin-color wave function for each channel is shown as follows. $\phi_{I_{z},s_{z} }^{B}$ represents the wave function of the $q^{3}$ cluster ($I_{z}$ and $s_{z}$ are the third component of the isospin and spin quantum numbers, $B$ is the corresponding baryon). Then we couple the two baryon wave functions by Clebsch-Gordan coefficients according to the total quantum number requirement, and we can obtain the total wave functions. 
There are seven channels for the $C=1,S=-1$ system:
\begin{align}
\left | \Lambda \Lambda _{c}   \right \rangle =&\phi _{0,\frac{1}{2} }^{\Lambda } \phi _{0,\frac{1}{2} }^{\Lambda _{c} } \nonumber \\
  \left | N\Xi _{c}  \right \rangle =&\sqrt{\frac{1}{2} } \left [ \phi _{\frac{1}{2},\frac{1}{2}  }^{p} \phi _{-\frac{1}{2} ,\frac{1}{2} }^{\Xi _{c} } -\phi _{-\frac{1}{2},\frac{1}{2}  }^{n} \phi _{\frac{1}{2} ,\frac{1}{2} }^{\Xi _{c} }\right ] \nonumber \\
  \left | N\Xi _{c}^{\prime }  \right \rangle =&\sqrt{\frac{1}{2} } \left [ \phi _{\frac{1}{2},\frac{1}{2}  }^{p} \phi _{-\frac{1}{2} ,\frac{1}{2} }^{\Xi _{c} ^{\prime} } -\phi _{-\frac{1}{2},\frac{1}{2}  }^{n} \phi _{\frac{1}{2} ,\frac{1}{2} }^{\Xi _{c}^{\prime } }\right ]  \nonumber \\
  \left | \Sigma \Sigma _{c} \right \rangle =&\sqrt{\frac{1}{3} } \left [ \phi _{1,\frac{1}{2} }^{\Sigma}\phi _{-1,\frac{1}{2} }^{\Sigma _{c} }-\phi _{0,\frac{1}{2} }^{\Sigma}\phi _{0,\frac{1}{2} }^{\Sigma _{c} }  +\phi _{-1,\frac{1}{2} }^{\Sigma}\phi _{1,\frac{1}{2} }^{\Sigma _{c} } \right ] \nonumber \\
  \left |\Sigma \Sigma _{c}^{*}  \right \rangle=&\frac{1}{2} \left [ \phi _{0,-\frac{1}{2} }^{\Sigma }\phi _{0,\frac{3}{2} }^{\Sigma _{c}^{*}}-\phi _{1,-\frac{1}{2} }^{\Sigma }\phi _{-1,\frac{3}{2} }^{\Sigma _{c}^{*}}-\phi _{-1,-\frac{1}{2} }^{\Sigma }\phi _{1,\frac{3}{2} }^{\Sigma _{c}^{*}}\right ]\nonumber \\
& -\sqrt{\frac{1}{12}} \left [ \phi _{0,\frac{1}{2} }^{\Sigma }\phi _{0,\frac{1}{2} }^{\Sigma _{c}^{*}}- \phi _{1,\frac{1}{2} }^{\Sigma }\phi _{-1,\frac{1}{2} }^{\Sigma _{c}^{*}}-\phi _{-1,\frac{1}{2} }^{\Sigma }\phi _{1,\frac{1}{2} }^{\Sigma _{c}^{*}}\right ] \nonumber \\
  \left |\Sigma^{*}  \Sigma _{c} \right \rangle=&\frac{1}{2} \left [ \phi _{1,\frac{3}{2} }^{\Sigma^{*} }\phi _{-1,-\frac{1}{2} }^{\Sigma _{c}}-\phi _{0,\frac{3}{2} }^{\Sigma^{*} }\phi _{0,-\frac{1}{2} }^{\Sigma _{c}}+\phi _{-1,\frac{3}{2} }^{\Sigma^{*} }\phi _{1,-\frac{1}{2} }^{\Sigma _{c}}\right ]\nonumber \\
& -\sqrt{\frac{1}{12}} \left [ \phi _{1,\frac{1}{2} }^{\Sigma ^{*}}\phi _{-1,\frac{1}{2} }^{\Sigma _{c}}- \phi _{0,\frac{1}{2} }^{\Sigma ^{*}}\phi _{0,\frac{1}{2} }^{\Sigma _{c}}-\phi _{-1,\frac{1}{2} }^{\Sigma^{*} }\phi _{1,\frac{1}{2} }^{\Sigma _{c}}\right ]   \nonumber \\
  \left |\Sigma^{*}  \Sigma _{c}^{*} \right \rangle=&\sqrt{\frac{1}{10} }  \left [ \phi _{1,\frac{3}{2} }^{\Sigma^{*} }\phi _{-1,-\frac{1}{2} }^{\Sigma _{c}^{*}}-\phi _{0,\frac{3}{2} }^{\Sigma^{*} }\phi _{0,-\frac{1}{2} }^{\Sigma _{c}^{*}}+\phi _{-1,\frac{3}{2} }^{\Sigma^{*} }\phi _{1,-\frac{1}{2} }^{\Sigma _{c}^{*}}\right.\nonumber \\
&\left.+\phi _{1,-\frac{1}{2} }^{\Sigma^{*} }\phi _{-1,\frac{3}{2} }^{\Sigma _{c}^{*}}-\phi _{0,-\frac{1}{2} }^{\Sigma^{*} }\phi _{0,\frac{3}{2} }^{\Sigma _{c}^{*}}+\phi _{-1,-\frac{1}{2} }^{\Sigma^{*} }\phi _{1,\frac{3}{2} }^{\Sigma _{c}^{*}}\right ]\nonumber \\
&-\sqrt{\frac{2}{15}} \left [ \phi _{1,\frac{1}{2} }^{\Sigma ^{*}}\phi _{-1,\frac{1}{2} }^{\Sigma _{c}^{*}}- \phi _{0,\frac{1}{2} }^{\Sigma ^{*}}\phi _{0,\frac{1}{2} }^{\Sigma _{c}^{*}}+\phi _{-1,\frac{1}{2} }^{\Sigma^{*} }\phi _{1,\frac{1}{2} }^{\Sigma _{c}^{*}}\right ]\nonumber
\end{align}

nine channels for the $C=1,S=-3$ system:
\begin{align}
\left | \Lambda \Omega _{c}\right \rangle=&\phi _{0,\frac{1}{2} }^{\Lambda } \phi _{0,\frac{1}{2} }^{\Omega _{c}}  \nonumber \\
\left | \Lambda \Omega _{c}^{*} \right \rangle =&\frac{1}{2}\phi _{0,\frac{1}{2} }^{\Lambda } \phi _{0,\frac{1}{2} }^{\Omega _{c}^{*}}-\sqrt{\frac{3}{4} }\phi _{0,-\frac{1}{2} }^{\Lambda } \phi _{0,\frac{3}{2} }^{\Omega _{c}^{*}} \nonumber \\
\left | \Lambda _{c}\Omega \right \rangle =&\frac{1}{2}\phi _{0,\frac{1}{2} }^{\Lambda _{c}} \phi _{0,\frac{1}{2} }^{\Omega }-\sqrt{\frac{3}{4} }\phi _{0,-\frac{1}{2} }^{\Lambda _{c}} \phi _{0,\frac{3}{2} }^{\Omega }\nonumber \\
\left |\Xi \Xi _{c}  \right \rangle =&\sqrt{\frac{1}{2} } \left [ \phi _{\frac{1}{2},\frac{1}{2}  }^{\Xi } \phi _{-\frac{1}{2},\frac{1}{2}  }^{\Xi _{c}} - \phi _{-\frac{1}{2},\frac{1}{2}  }^{\Xi } \phi _{\frac{1}{2},\frac{1}{2}}^{\Xi _{c}} \right ]\nonumber \\
\left |\Xi \Xi _{c}^{\prime }  \right \rangle =&\sqrt{\frac{1}{2} } \left [ \phi _{\frac{1}{2},\frac{1}{2}  }^{\Xi } \phi _{-\frac{1}{2},\frac{1}{2}  }^{\Xi _{c}^{\prime }} - \phi _{-\frac{1}{2},\frac{1}{2}  }^{\Xi } \phi _{\frac{1}{2},\frac{1}{2}  }^{\Xi _{c}^{\prime }} \right ] \nonumber \\
\left |\Xi \Xi _{c}^{*}\right \rangle =&\sqrt{\frac{3}{8}}\left [ \phi _{-\frac{1}{2},-\frac{1}{2}  }^{\Xi } \phi _{\frac{1}{2},\frac{3}{2}  }^{\Xi _{c}^{* }} - \phi _{\frac{1}{2},-\frac{1}{2}  }^{\Xi } \phi _{-\frac{1}{2},\frac{3}{2}  }^{\Xi _{c}^{* }} \right ]\nonumber \\
&-\sqrt{\frac{1}{8} }\left [\phi _{-\frac{1}{2},\frac{1}{2}  }^{\Xi } \phi _{\frac{1}{2},\frac{1}{2}  }^{\Xi _{c}^{* }} - \phi _{\frac{1}{2},\frac{1}{2}  }^{\Xi } \phi _{-\frac{1}{2},\frac{1}{2}  }^{\Xi _{c}^{* }}  \right ] \nonumber \\
\left |\Xi^{*} \Xi _{c}\right \rangle =&\sqrt{\frac{3}{8}}\left [ \phi _{\frac{1}{2},\frac{3}{2}  }^{\Xi ^{* }} \phi _{-\frac{1}{2},-\frac{1}{2}  }^{\Xi _{c}} - \phi _{-\frac{1}{2},\frac{3}{2}  }^{\Xi ^{* }} \phi _{\frac{1}{2},-\frac{1}{2}  }^{\Xi _{c}} \right ]\nonumber \\
&-\sqrt{\frac{1}{8} }\left [\phi _{\frac{1}{2},\frac{1}{2}  }^{\Xi^{* } } \phi _{-\frac{1}{2},\frac{1}{2}  }^{\Xi _{c}} - \phi _{-\frac{1}{2},\frac{1}{2}  }^{\Xi ^{* }} \phi _{\frac{1}{2},\frac{1}{2}  }^{\Xi _{c}}  \right ] \nonumber \\
\left |\Xi^{*} \Xi _{c}^{\prime }\right \rangle =&\sqrt{\frac{3}{8}}\left [ \phi _{\frac{1}{2},\frac{3}{2}  }^{\Xi ^{* }} \phi _{-\frac{1}{2},-\frac{1}{2}  }^{\Xi _{c}^{\prime }} - \phi _{-\frac{1}{2},\frac{3}{2}  }^{\Xi ^{* }} \phi _{\frac{1}{2},-\frac{1}{2}  }^{\Xi _{c}^{\prime }} \right ]\nonumber \\
&-\sqrt{\frac{1}{8} }\left [\phi _{\frac{1}{2},\frac{1}{2}  }^{\Xi^{* } } \phi _{-\frac{1}{2},\frac{1}{2}  }^{\Xi _{c}^{\prime }} - \phi _{-\frac{1}{2},\frac{1}{2}  }^{\Xi ^{* }} \phi _{\frac{1}{2},\frac{1}{2}  }^{\Xi _{c}^{\prime }}  \right ]  \nonumber \\
\left |\Xi^{*} \Xi _{c}^{*}\right \rangle =&\sqrt{\frac{3}{20}}\left [ \phi _{\frac{1}{2},\frac{3}{2}  }^{\Xi ^{* }} \phi _{-\frac{1}{2},-\frac{1}{2}  }^{\Xi _{c}^{*}} - \phi _{-\frac{1}{2},\frac{3}{2}  }^{\Xi ^{* }} \phi _{\frac{1}{2},-\frac{1}{2}  }^{\Xi _{c}^{*}} \right.\nonumber \\
&\left.+\phi _{\frac{1}{2},-\frac{1}{2}  }^{\Xi ^{* }} \phi _{-\frac{1}{2},\frac{3}{2}  }^{\Xi _{c}^{*}}-\phi _{-\frac{1}{2},-\frac{1}{2}  }^{\Xi ^{* }} \phi _{\frac{1}{2},\frac{3}{2}  }^{\Xi _{c}^{*}}\right ]\nonumber \\
&-\sqrt{\frac{3}{10} }\left [\phi _{\frac{1}{2},\frac{1}{2}  }^{\Xi ^{* }} \phi _{-\frac{1}{2},\frac{1}{2}  }^{\Xi _{c}^{*}}-\phi _{-\frac{1}{2},\frac{1}{2}  }^{\Xi ^{* }} \phi _{\frac{1}{2},\frac{1}{2}  }^{\Xi _{c}^{*}}\right ] \nonumber
\end{align}

and two channels for the $C=1,S=-5$ system:
\begin{align}
\left | \Omega \Omega _{c}\right \rangle=&\sqrt{-\frac{3}{4} } \phi _{0,-\frac{1}{2} }^{\Omega_{c}}\phi _{0,\frac{3}{2} }^{\Omega}+\sqrt{\frac{1}{4} }\phi _{0,\frac{1}{2} }^{\Omega_{c}}\phi _{0,\frac{1}{2} }^{\Omega}  \nonumber \\
\left | \Omega \Omega _{c}^{*}  \right \rangle =&\sqrt{\frac{3}{10} }\left [ \phi _{0,\frac{3}{2} }^{\Omega} \phi _{0,-\frac{1}{2} }^{\Omega_{c}^{*} } +\phi _{0,-\frac{1}{2} }^{\Omega} \phi _{0,\frac{3}{2} }^{\Omega_{c}^{*} }\right ] \nonumber \\
&-\sqrt{\frac{2}{5} }\phi _{0,\frac{1}{2}}^{\Omega }\phi _{0,\frac{1}{2} }^{\Omega _{c} ^{*}} \nonumber
\end{align}

where the expression of $\phi_{I_{z},s_{z} }^{B}$ is shown as follows:
\begin{align}
\phi _{0,\frac{1}{2}}^{\Lambda}=& \sqrt{\frac{1}{2} }\left (  \chi _{0,0}^{f1}\chi _{\frac{1}{2},\frac{1}{2}}^{\sigma 1} +\chi _{0,0}^{f2}\chi _{\frac{1}{2},\frac{1}{2}}^{\sigma 2} \right )\chi ^{c}\nonumber \\
\phi _{0,\frac{1}{2}}^{\Lambda_{c}}=& \sqrt{\frac{1}{2}}\left (  \chi _{0,0}^{f3}\chi _{\frac{1}{2},\frac{1}{2}}^{\sigma 1} +\chi _{0,0}^{f4}\chi _{\frac{1}{2},\frac{1}{2}}^{\sigma 2} \right )\chi ^{c}\nonumber \\
\phi _{\frac{1}{2},\frac{1}{2}}^{p}=& \sqrt{\frac{1}{2}}\left (  \chi _{\frac{1}{2},\frac{1}{2}}^{f1}\chi _{\frac{1}{2},\frac{1}{2}}^{\sigma 1} +\chi _{\frac{1}{2},\frac{1}{2}}^{f2}\chi _{\frac{1}{2},\frac{1}{2}}^{\sigma 2} \right )\chi ^{c}\nonumber \\
\phi _{-\frac{1}{2},\frac{1}{2}}^{\Xi_{c}}=& \sqrt{\frac{1}{2}}\left (  \chi _{\frac{1}{2},-\frac{1}{2}}^{f1}\chi _{\frac{1}{2},\frac{1}{2}}^{\sigma 1} +\chi _{\frac{1}{2},-\frac{1}{2}}^{f2}\chi _{\frac{1}{2},\frac{1}{2}}^{\sigma 2} \right )\chi ^{c}\nonumber \\
\phi _{-\frac{1}{2},\frac{1}{2}}^{n}=& \sqrt{\frac{1}{2}}\left (  \chi _{\frac{1}{2},-\frac{1}{2}}^{f3}\chi _{\frac{1}{2},\frac{1}{2}}^{\sigma 1} +\chi _{\frac{1}{2},-\frac{1}{2}}^{f4}\chi _{\frac{1}{2},\frac{1}{2}}^{\sigma 2} \right )\chi ^{c}\nonumber \\
\phi _{\frac{1}{2},\frac{1}{2}}^{\Xi_{c}}=& \sqrt{\frac{1}{2}}\left (  \chi _{\frac{1}{2},\frac{1}{2}}^{f3}\chi _{\frac{1}{2},\frac{1}{2}}^{\sigma 1} +\chi _{\frac{1}{2},\frac{1}{2}}^{f4}\chi _{\frac{1}{2},\frac{1}{2}}^{\sigma 2} \right )\chi ^{c}\nonumber \\
\phi _{-\frac{1}{2},\frac{1}{2}}^{\Xi_{c}^{\prime}}=& \sqrt{\frac{1}{2}}\left (  \chi _{\frac{1}{2},-\frac{1}{2}}^{f5}\chi _{\frac{1}{2},\frac{1}{2}}^{\sigma 1} +\chi _{\frac{1}{2},-\frac{1}{2}}^{f6}\chi _{\frac{1}{2},\frac{1}{2}}^{\sigma 2} \right )\chi ^{c}\nonumber \\
\phi _{\frac{1}{2},\frac{1}{2}}^{\Xi_{c}^{\prime}}=& \sqrt{\frac{1}{2}}\left (  \chi _{\frac{1}{2},\frac{1}{2}}^{f5}\chi _{\frac{1}{2},\frac{1}{2}}^{\sigma 1} +\chi _{\frac{1}{2},\frac{1}{2}}^{f6}\chi _{\frac{1}{2},\frac{1}{2}}^{\sigma 2} \right )\chi ^{c}\nonumber \\
\phi _{1,\frac{1}{2}}^{\Sigma}=& \sqrt{\frac{1}{2}}\left (  \chi _{1,1}^{f1}\chi _{\frac{1}{2},\frac{1}{2}}^{\sigma 1} +\chi _{1,1}^{f2}\chi _{\frac{1}{2},\frac{1}{2}}^{\sigma 2} \right )\chi ^{c}\nonumber \\
\phi _{-1,\frac{1}{2}}^{\Sigma_{c}}=& \sqrt{\frac{1}{2}}\left (  \chi _{1,-1}^{f1}\chi _{\frac{1}{2},\frac{1}{2}}^{\sigma 1} +\chi _{1,-1}^{f2}\chi _{\frac{1}{2},\frac{1}{2}}^{\sigma 2} \right )\chi ^{c}\nonumber \\
\phi _{0,\frac{1}{2}}^{\Sigma}=& \sqrt{\frac{1}{2}}\left (  \chi _{1,0}^{f1}\chi _{\frac{1}{2},\frac{1}{2}}^{\sigma 1} +\chi _{1,0}^{f2}\chi _{\frac{1}{2},\frac{1}{2}}^{\sigma 2} \right )\chi ^{c}\nonumber \\
\phi _{0,\frac{1}{2}}^{\Sigma_{c}}=& \sqrt{\frac{1}{2}}\left (  \chi _{1,0}^{f3}\chi _{\frac{1}{2},\frac{1}{2}}^{\sigma 1} +\chi _{1,0}^{f4}\chi _{\frac{1}{2},\frac{1}{2}}^{\sigma 2} \right )\chi ^{c}\nonumber \\
\phi _{-1,\frac{1}{2}}^{\Sigma}=& \sqrt{\frac{1}{2}}\left (  \chi _{1,-1}^{f3}\chi _{\frac{1}{2},\frac{1}{2}}^{\sigma 1} +\chi _{1,-1}^{f4}\chi _{\frac{1}{2},\frac{1}{2}}^{\sigma 2} \right )\chi ^{c}\nonumber \\
\phi _{1,\frac{1}{2}}^{\Sigma_{c}}=& \sqrt{\frac{1}{2}}\left (  \chi _{1,1}^{f3}\chi _{\frac{1}{2},\frac{1}{2}}^{\sigma 1} +\chi _{1,1}^{f4}\chi _{\frac{1}{2},\frac{1}{2}}^{\sigma 2} \right )\chi ^{c}\nonumber \\
\phi _{0,-\frac{1}{2}}^{\Sigma}=& \sqrt{\frac{1}{2}}\left (  \chi _{1,0}^{f1}\chi _{\frac{1}{2},-\frac{1}{2}}^{\sigma 1} +\chi _{1,0}^{f2}\chi _{\frac{1}{2},-\frac{1}{2}}^{\sigma 2} \right )\chi ^{c}\nonumber \\
\phi _{1,-\frac{1}{2}}^{\Sigma}=& \sqrt{\frac{1}{2}}\left (  \chi _{1,1}^{f1}\chi _{\frac{1}{2},-\frac{1}{2}}^{\sigma 1} +\chi _{1,1}^{f2}\chi _{\frac{1}{2},-\frac{1}{2}}^{\sigma 2} \right )\chi ^{c}\nonumber \\
\phi _{-1,-\frac{1}{2}}^{\Sigma}=& \sqrt{\frac{1}{2}}\left (  \chi _{1,-1}^{f3}\chi _{\frac{1}{2},-\frac{1}{2}}^{\sigma 1} +\chi _{1,-1}^{f4}\chi _{\frac{1}{2},-\frac{1}{2}}^{\sigma 2} \right )\chi ^{c}\nonumber \\
\phi _{0,\frac{1}{2}}^{\Omega_{c}}=& \sqrt{\frac{1}{2} }\left (  \chi _{0,0}^{f5}\chi _{\frac{1}{2},\frac{1}{2}}^{\sigma 1} +\chi _{0,0}^{f6}\chi _{\frac{1}{2},\frac{1}{2}}^{\sigma 2} \right )\chi ^{c}\nonumber \\
\phi _{0,-\frac{1}{2}}^{\Lambda}=& \sqrt{\frac{1}{2} }\left (  \chi _{0,0}^{f1}\chi _{\frac{1}{2},-\frac{1}{2}}^{\sigma 1} +\chi _{0,0}^{f2}\chi _{\frac{1}{2},-\frac{1}{2}}^{\sigma 2} \right )\chi ^{c}\nonumber \\
\phi _{\frac{1}{2},\frac{1}{2}}^{\Xi}=& \sqrt{\frac{1}{2} }\left (  \chi _{\frac{1}{2},\frac{1}{2}}^{f7}\chi _{\frac{1}{2},\frac{1}{2}}^{\sigma 1} +\chi _{\frac{1}{2},\frac{1}{2}}^{f8}\chi _{\frac{1}{2},\frac{1}{2}}^{\sigma 2} \right )\chi ^{c}\nonumber \\
\phi _{-\frac{1}{2},\frac{1}{2}}^{\Xi}=& \sqrt{\frac{1}{2} }\left (  \chi _{\frac{1}{2},-\frac{1}{2}}^{f7}\chi _{\frac{1}{2},\frac{1}{2}}^{\sigma 1} +\chi _{\frac{1}{2},-\frac{1}{2}}^{f8}\chi _{\frac{1}{2},\frac{1}{2}}^{\sigma 2} \right )\chi ^{c}\nonumber \\
\phi _{-\frac{1}{2},-\frac{1}{2}}^{\Xi}=& \sqrt{\frac{1}{2} }\left (  \chi _{\frac{1}{2},-\frac{1}{2}}^{f7}\chi _{\frac{1}{2},-\frac{1}{2}}^{\sigma 1} +\chi _{\frac{1}{2},-\frac{1}{2}}^{f8}\chi _{\frac{1}{2},-\frac{1}{2}}^{\sigma 2} \right )\chi ^{c}\nonumber \\
\phi _{\frac{1}{2},-\frac{1}{2}}^{\Xi}=& \sqrt{\frac{1}{2} }\left (  \chi _{\frac{1}{2},\frac{1}{2}}^{f7}\chi _{\frac{1}{2},-\frac{1}{2}}^{\sigma 1} +\chi _{\frac{1}{2},\frac{1}{2}}^{f8}\chi _{\frac{1}{2},-\frac{1}{2}}^{\sigma 2} \right )\chi ^{c}\nonumber \\
\phi _{-\frac{1}{2},-\frac{1}{2}}^{\Xi_{c}^{\prime}}=& \sqrt{\frac{1}{2} }\left (  \chi _{\frac{1}{2},-\frac{1}{2}}^{f5}\chi _{\frac{1}{2},-\frac{1}{2}}^{\sigma 1} +\chi _{\frac{1}{2},-\frac{1}{2}}^{f6}\chi _{\frac{1}{2},-\frac{1}{2}}^{\sigma 2} \right )\chi ^{c}\nonumber \\
\phi _{\frac{1}{2},-\frac{1}{2}}^{\Xi_{c}^{\prime}}=& \sqrt{\frac{1}{2} }\left (  \chi _{\frac{1}{2},\frac{1}{2}}^{f5}\chi _{\frac{1}{2},-\frac{1}{2}}^{\sigma 1} +\chi _{\frac{1}{2},\frac{1}{2}}^{f6}\chi _{\frac{1}{2},-\frac{1}{2}}^{\sigma 2} \right )\chi ^{c}\nonumber 
\end{align}
\begin{align}
\phi _{0,-\frac{1}{2}}^{\Omega_{c}}=& \sqrt{\frac{1}{2} }\left (  \chi _{0,0}^{f5}\chi _{\frac{1}{2},-\frac{1}{2}}^{\sigma 1} +\chi _{0,0}^{f6}\chi _{\frac{1}{2},-\frac{1}{2}}^{\sigma 2} \right )\chi ^{c}\nonumber
\end{align}

\begin{align}
\phi_{0,\frac{3}{2}}^{\Sigma_{c}^*}=&\chi_{1,0}^{f5}\chi_
{\frac{3}{2},\frac{3}{2}}^{\sigma}\chi^{c}, ~~~~~~~~~~~~\phi_{-1,\frac{3}{2}}^{\Sigma_{c}^*}=\chi_{1,-1}^{f5}
\chi_{\frac{3}{2},\frac{3}{2}}^{\sigma}\chi^{c}\nonumber \\
\phi_{1,\frac{3}{2}}^{\Sigma_{c}^*}=&\chi_{1,1}^{f5}\chi_
{\frac{3}{2},\frac{3}{2}}^{\sigma}\chi^{c}, ~~~~~~~~~~~~\phi_{0,\frac{1}{2}}^{\Sigma_{c}^*}=\chi_{1,0}^{f5}
\chi_{\frac{3}{2},\frac{1}{2}}^{\sigma}\chi^{c}\nonumber \\
\phi_{-1,\frac{1}{2}}^{\Sigma_{c}^*}=&\chi_{1,-1}^{f5}\chi_
{\frac{3}{2},\frac{1}{2}}^{\sigma}\chi^{c}, ~~~~~~~~~~~~\phi_{1,\frac{1}{2}}^{\Sigma_{c}^*}=\chi_{1,1}^{f5}
\chi_{\frac{3}{2},\frac{1}{2}}^{\sigma}\chi^{c}\nonumber \\
\phi_{1,\frac{3}{2}}^{\Sigma^*}=&\chi_{1,1}^{f6}\chi_
{\frac{3}{2},\frac{3}{2}}^{\sigma}\chi^{c}, ~~~~~~~~~~~~\phi_{0,\frac{3}{2}}^{\Sigma^*}=\chi_{1,0}^{f6}
\chi_{\frac{3}{2},\frac{3}{2}}^{\sigma}\chi^{c}\nonumber \\
\phi_{-1,\frac{3}{2}}^{\Sigma^*}=&\chi_{1,-1}^{f6}\chi_
{\frac{3}{2},\frac{3}{2}}^{\sigma}\chi^{c}, ~~~~~~~~~~~~\phi_{1,\frac{1}{2}}^{\Sigma^*}=\chi_{1,1}^{f6}
\chi_{\frac{3}{2},\frac{1}{2}}^{\sigma}\chi^{c}\nonumber \\
\phi_{0,\frac{1}{2}}^{\Sigma^*}=&\chi_{1,0}^{f6}\chi_
{\frac{3}{2},\frac{1}{2}}^{\sigma}\chi^{c}, ~~~~~~~~~~~~\phi_{-1,\frac{1}{2}}^{\Sigma^*}=\chi_{1,-1}^{f6}
\chi_{\frac{3}{2},\frac{1}{2}}^{\sigma}\chi^{c}\nonumber \\
\phi_{-1,-\frac{1}{2}}^{\Sigma_{c}^*}=&\chi_{1,-1}^{f5}\chi_
{\frac{3}{2},-\frac{1}{2}}^{\sigma}\chi^{c}, ~~~~~~~~~~~~\phi_{0,-\frac{1}{2}}^{\Sigma_{c}^*}=\chi_{1,0}^{f5}
\chi_{\frac{3}{2},-\frac{1}{2}}^{\sigma}\chi^{c}\nonumber \\
\phi_{1,-\frac{1}{2}}^{\Sigma_{c}^*}=&\chi_{1,1}^{f5}\chi_
{\frac{3}{2},-\frac{1}{2}}^{\sigma}\chi^{c}, ~~~~~~~~~~~~\phi_{1,-\frac{1}{2}}^{\Sigma^*}=\chi_{1,1}^{f6}
\chi_{\frac{3}{2},-\frac{1}{2}}^{\sigma}\chi^{c}\nonumber \\
\phi_{0,-\frac{1}{2}}^{\Sigma_{c}^*}=&\chi_{1,0}^{f6}\chi_
{\frac{3}{2},-\frac{1}{2}}^{\sigma}\chi^{c}, ~~~~~~~~~~~~\phi_{-1,-\frac{1}{2}}^{\Sigma^*}=\chi_{1,-1}^{f6}
\chi_{\frac{3}{2},-\frac{1}{2}}^{\sigma}\chi^{c}\nonumber \\
\phi_{0,\frac{3}{2}}^{\Omega_{c}^*}=&\chi_{0,0}^{f7}\chi_
{\frac{3}{2},\frac{3}{2}}^{\sigma}\chi^{c}, ~~~~~~~~~~~~\phi_{0,\frac{1}{2}}^{\Omega_{c}^*}=\chi_{0,0}^{f7}
\chi_{\frac{3}{2},\frac{1}{2}}^{\sigma}\chi^{c}\nonumber \\
\phi_{0,\frac{1}{2}}^{\Omega}=&\chi_{0,0}^{f8}\chi_
{\frac{3}{2},\frac{1}{2}}^{\sigma}\chi^{c}, ~~~~~~~~~~~~\phi_{0,\frac{3}{2}}^{\Omega}=\chi_{0,0}^{f8}
\chi_{\frac{3}{2},\frac{3}{2}}^{\sigma}\chi^{c}\nonumber \\
\phi_{\frac{1}{2},\frac{3}{2}}^{\Xi_{c}^*}=&\chi_
{\frac{1}{2},\frac{1}{2}}^{f9}\chi_{\frac{3}{2},
\frac{3}{2}}^{\sigma}\chi^{c}, ~~~~~~~~~~~~\phi_{-\frac{1}{2},\frac{3}{2}}^{\Xi_{c}^*}=\chi_
{\frac{1}{2},-\frac{1}{2}}^{f9}\chi_{\frac{3}{2},
\frac{3}{2}}^{\sigma}\chi^{c}\nonumber\\
\phi_{\frac{1}{2},\frac{1}{2}}^{\Xi_{c}^*}=&\chi_
{\frac{1}{2},\frac{1}{2}}^{f9}\chi_{\frac{3}{2},
\frac{1}{2}}^{\sigma}\chi^{c}, ~~~~~~~~~~~~\phi_{-\frac{1}{2},\frac{1}{2}}^{\Xi_{c}^*}=\chi_
{\frac{1}{2},-\frac{1}{2}}^{f9}\chi_{\frac{3}{2},
\frac{1}{2}}^{\sigma}\chi^{c}\nonumber\\
\phi_{\frac{1}{2},\frac{3}{2}}^{\Xi^*}=&\chi_
{\frac{1}{2},\frac{1}{2}}^{f10}\chi_{\frac{3}{2},
\frac{3}{2}}^{\sigma}\chi^{c}, ~~~~~~~~~~~~\phi_{-\frac{1}{2},\frac{3}{2}}^{\Xi^*}=\chi_
{\frac{1}{2},-\frac{1}{2}}^{f10}\chi_{\frac{3}{2},
\frac{3}{2}}^{\sigma}\chi^{c}\nonumber\\
\phi_{\frac{1}{2},\frac{1}{2}}^{\Xi^*}=&\chi_
{\frac{1}{2},\frac{1}{2}}^{f10}\chi_{\frac{3}{2},
\frac{1}{2}}^{\sigma}\chi^{c}, ~~~~~~~~~~~~\phi_{-\frac{1}{2},\frac{1}{2}}^{\Xi^*}=\chi_
{\frac{1}{2},-\frac{1}{2}}^{f10}\chi_{\frac{3}{2},
\frac{1}{2}}^{\sigma}\chi^{c}\nonumber\\
\phi_{-\frac{1}{2},-\frac{1}{2}}^{\Xi_{c}^*}=&\chi_
{\frac{1}{2},-\frac{1}{2}}^{f9}\chi_{\frac{3}{2},
-\frac{1}{2}}^{\sigma}\chi^{c}, ~~~~~~~~~~~~\phi_{\frac{1}{2},-\frac{1}{2}}^{\Xi_{c}^*}=\chi_
{\frac{1}{2},\frac{1}{2}}^{f9}\chi_{\frac{3}{2},
-\frac{1}{2}}^{\sigma}\chi^{c}\nonumber\\
\phi_{\frac{1}{2},-\frac{1}{2}}^{\Xi^*}=&\chi_
{\frac{1}{2},\frac{1}{2}}^{f10}\chi_{\frac{3}{2},
-\frac{1}{2}}^{\sigma}\chi^{c}, ~~~~~~~~~~~~\phi_{-\frac{1}{2},-\frac{1}{2}}^{\Xi^*}=\chi_
{\frac{1}{2},-\frac{1}{2}}^{f10}\chi_{\frac{3}{2},
-\frac{1}{2}}^{\sigma}\chi^{c}\nonumber\\
\phi_{0,-\frac{1}{2}}^{\Omega}=&\chi_{0,0}^{f8}\chi_
{\frac{3}{2},-\frac{1}{2}}^{\sigma}\chi^{c}, ~~~~~~~~~~~~\phi_{0,\frac{3}{2}}^{\Omega_{c}^*}=\chi_{0,0}^{f7}
\chi_{\frac{3}{2},-\frac{1}{2}}^{\sigma}\chi^{c}\nonumber
\end{align}

\end{document}